\documentclass[twocolumn]{aastex63_ionthinspace}

\usepackage{xcolor}
\usepackage{amsmath}
\usepackage{affils}

\newcommand{\HI}{\ion{H}{1}}

\newcommand{\hi}[0]{\HI{}}

\newcommand{\reff}[0]{${\rm R_{\rm eff}}$}
\newcommand{\sersic}[0]{S\'ersic}

\newcommand{\thh}[0]{$^{\rm th}$}

\newcommand{\galex}[0]{\textit{GALEX}}
\newcommand{\hsc}[0]{HSC}

\newcommand{\rrr}[1]{\textcolor{black}{#1}}
\newcommand{\rrrtwo}[1]{\textcolor{black}{#1}}

\newcommand{\av}[0]{$ A_V$}
\newcommand{\step}[1]{step \textsf{#1}}

\newcommand{\hscdr}[0]{S21A}

\newcommand{\sigsfr}[0]{$\Sigma_{\rm SFR}$}

\defcitealias{papertwo}{Paper II}

\begin{document}
\shortauthors{Kado-Fong et al.}

\title{Ultra-Diffuse Galaxies as Extreme Star-forming Environments I: Mapping Star Formation in \HI{}-Rich UDGs}

\DeclareAffil{princeton}{Department of Astrophysical Sciences, Princeton University,Princeton, NJ 08544, USA}

\author[0000-0002-0332-177X]{Erin Kado-Fong}
\affiliation{Department of Astrophysical Sciences, Princeton University,Princeton, NJ 08544, USA}
\author[0000-0002-5612-3427]{Jenny E. Greene}
\affiliation{Department of Astrophysical Sciences, Princeton University,Princeton, NJ 08544, USA}
\author{Song Huang}
\affiliation{Department of Astronomy, Tsinghua University, Beijing 100084, China}
\author{Andy Goulding}
\affiliation{Department of Astrophysical Sciences, Princeton University,Princeton, NJ 08544, USA}
\correspondingauthor{Erin Kado-Fong} 
\email{kadofong@princeton.edu}
  
  \date{\today}

\begin{abstract}
Ultra-Diffuse Galaxies are both extreme products of galaxy evolution and extreme environments
in which to test our understanding of star formation. In this work, we contrast the 
spatially resolved star formation activity of a sample of
22 \HI-selected UDGs and 35 low-mass galaxies from the NASA Sloan Atlas \rrr{(NSA)} within 120 Mpc. We employ
a new joint SED fitting method to compute star formation rate and stellar mass surface density
maps that leverage the high spatial resolution optical imaging data of the Hyper Suprime-Cam Subaru
Strategic Program (HSC-SSP) and the UV coverage of \galex{}, along with \HI{} radial profiles estimated
from a subset of galaxies that have spatially resolved \HI{} maps. We find that the UDGs have 
low star formation efficiencies as a function of their atomic gas down to scales of 500 pc. 
 We additionally find that the stellar \rrr{mass-weighted} sizes of our UDG sample
are unremarkable when considered as a function of their \HI{} mass \rrr{-- their stellar sizes are
comparable to the NSA dwarfs at fixed \HI{} mass}. This 
is a natural result in
the picture where UDGs are forming stars normally, but at low efficiencies. We compare our results
to predictions from contemporary models of galaxy formation, and find in particular that our 
observations are difficult to reproduce in models where UDGs undergo stellar expansion due to 
vigorous star formation feedback \rrr{should bursty star formation be required down to $z=0$}. 
\end{abstract}

\section{Introduction}

\begin{figure*}[htb]
\centering     
\includegraphics[width=\linewidth]{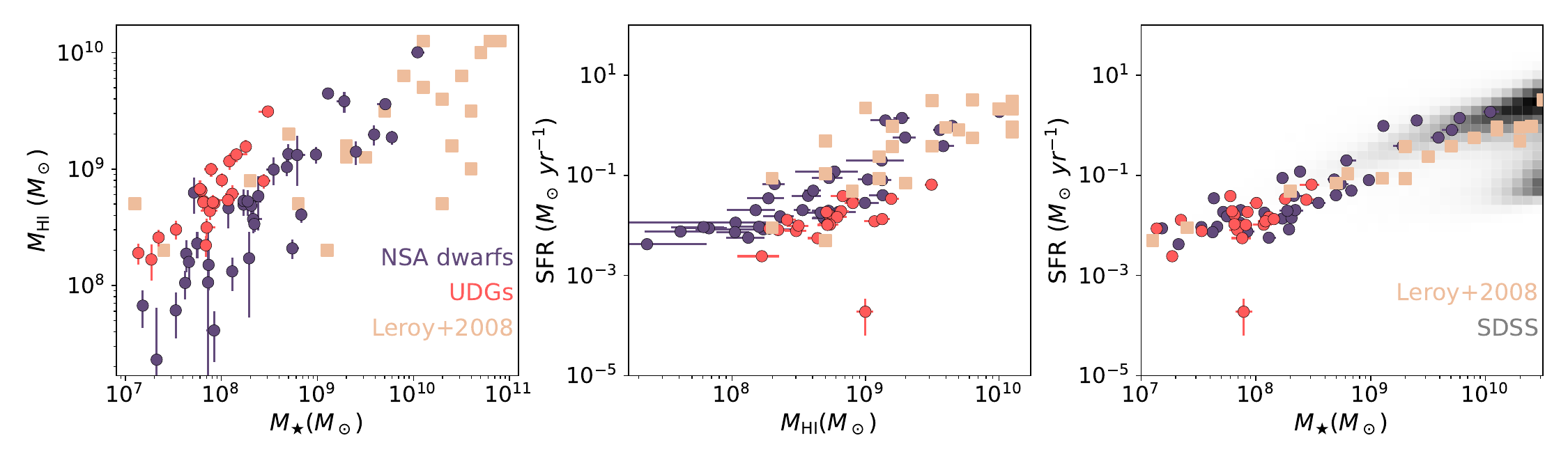}
\caption{ 
    A comparison of the integrated properties of the samples derived in this work
    against the directly measured result of \cite{leroy2008} (beige points). In all
    panels, the red points show UDGs and the purple points show NSA dwarfs. From left,
    we show the relationship between stellar mass and \HI{} mass, the relationship between
    \HI{} mass and SFR, and the star-forming main sequence (SFMS). In the SFMS panel, 
    we also show the results of the SDSS DR7 MPA-JHU added-value catalogs in greyscale
    \citep{kauffmann2003, brinchmann2004, salim2007}. We find a good agreement between
    our results and those from the literature. Although the UDGs have high \HI{} masses
    for their stellar mass (left panel), they have relatively little SFR for their \HI{}
    mass (middle panel).
    }
\label{f:gascomparison}
\end{figure*}

\begin{figure}[htb]
\centering     
\includegraphics[width=\linewidth]{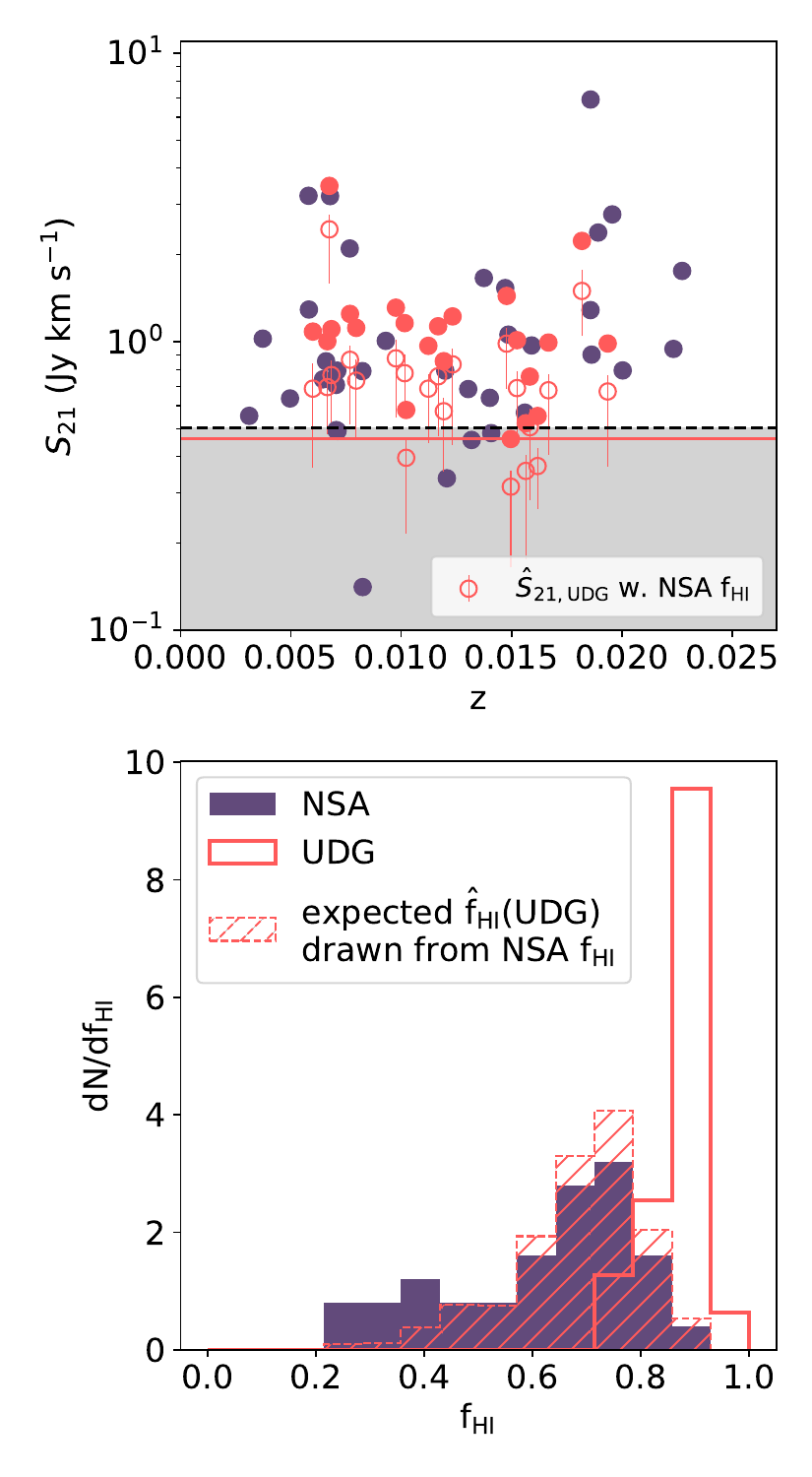}
\caption{ 
    \textit{Top:} 21 cm line flux ($S_{21}$) versus redshift for the NSA (purple) and
    UDG (red) samples. The nominal ALFALFA flux limit is shown by the dashed black line,
    while the minimum $S_{21}$ of the UDG sample is shown in red. The unfilled symbols show the expected 
    $S_{21}$ of each of the galaxies in the UDG sample \textit{if} we assume that the
    distribution over \HI{} fraction of the underlying UDG population is identical
    to that of the NSA dwarfs.
    \textit{Bottom:} the distribution over observed \HI{} fractions for the NSA (filled
    purple histogram) and UDG (unfilled red histogram) samples. The hatched red histogram
    shows the distribution of \HI{} fractions that we would \textit{expect} to see from
    the UDG sample if the distribution over \HI{} fractions is identical between the
    UDG population and NSA sample. We thus conclude that the UDG sample is inconsistent with 
    having the same distribution over $\rm f_{\rm HI}$ as the NSA sample.}
\label{f:fhi_detection}
\end{figure}

Ultra-diffuse galaxies (UDGs) are
dwarf galaxies that are characterized by large stellar sizes 
(\reff$>1.5$ kpc) and low surface brightnesses ($\mu_{0,g} > 24$ mag arcsec$^{-2}$, though
exact definitions vary; see \citealt{vannest2022} for an overview). These
diffuse systems
are extreme
in two senses: they lie far above the
mass-size relation, making them extreme products of galaxy evolution, but
they are also extreme environments for star formation due to their
low stellar surface densities.

UDGs as extreme products of galaxy evolution is a well-trodden topic both observationally 
\citep[see, e.g.][]{sandage1984, mcgaugh1995, dalcanton1997, vandokkum2015, beasley2016, beasley2016b, peng2016, yagi2016, leisman2017, vandokkum2018, danieli2019, janowiecki2019, vandokkum2019, danieli2021, gault2021} and theoretically \citep[e.g.][]{amorisco2016, dicintio2017, chan2018, greco2018a, greco2018b, jiang2019, liao2019, wright2021, vannest2022, zaritsky2022}. 
However, the formation of the UDG population remains an open question. 
Several mechanisms
have been proposed to form UDGs in the field,
including stellar expansion from star formation 
feedback \citep{elbadry2016, dicintio2017, chan2018}, early mergers that trigger 
radial expansion of star formation \citep{wright2021}, and preferential UDG formation
in the high-spin tail of the halo distribution \citep{dalcanton1997, liao2019}. 

It is observationally challenging to measure robust distances to 
UDGs due to their low surface brightnesses; as a result, the majority of UDGs with
known distances are in clusters or groups. To understand the full breadth of the UDG 
population, of course, it is necessary to sample this population across a range of
environments. 
Observational works continue to place new constraints on such mechanisms 
\citep[see, e.g.][]{leisman2017, greco2018a, greco2018b, janowiecki2019, martinnavarro2019, gault2021, kadofong2021a, greene2022, villaume2022}, but as of yet there has not been a clear
convergence in the literature towards one path -- of course, it is also not
necessarily the case that all UDGs are formed via the same mechanism.
Towards this end, there has been a push to use \HI{} to detect 
and/or measure distances to UDGs \citep{leisman2017, janowiecki2019, karunakaran2020, gault2021} -- 
this allows for distance determinations in a sample of UDGs that tend towards
low density environments.

It has been shown that although \HI-selected UDGs lie
on the star-forming main sequence and \HI{} mass-size relation
\citep{gault2021}, they tend to have high \HI{} masses \citep{leisman2017}
-- that is, the UDGs are, in an integrated sense, low in their star formation efficiency
as a function of atomic gas, or SFE(\HI). 
This in and of itself is not surprising -- it has been 
demonstrated observationally
that stellar mass surface density affects the position of dwarf
galaxies with respect to the 
Kennicutt-Schmidt relation \citep{schmidt1959, kennicutt1989} in that
lower stellar mass surface density dwarfs lie below the main relation
\citep{delosreyes2019}.
What has not yet been ascertained, however, is whether the
apparently low SFE(\HI) of the UDGs is a product of their selection, the distribution of their
star formation, or a true physical suppression of SFE(\HI). 

In order to understand the origin of the
low SFE(\HI) of the UDGs,
we introduce a joint
SED fitting method that utilizes the spatial resolution of the optical Hyper Suprime-Cam Subaru
Strategic Program (HSC-SSP) and the crucial UV coverage of \galex{}. We use this method to derive
stellar mass and SFR maps of a sample of \HI-selected UDGs from \cite{janowiecki2019} 
and a reference sample of \HI-detected galaxies from the NASA Sloan Atlas \citep{bradford2015}. We then use these maps,
along with estimated \HI{} radial profiles, to probe the star formation efficiency of these
dwarfs down to 500 pc scales. In this work, our main focus will be on 
characterizing the properties of the \HI-selected UDGs with respect to the 
``normal'' NSA dwarf sample through a lens aimed largely at galaxy evolution. 
In \cite{papertwo} (hereafter \citetalias{papertwo}) we will
shift our focus towards understanding star formation in \HI-rich UDGs, with a 
particular focus on the role of UDG structure in the pressure-regulated,
feedback-modulated model of star formation \citep{ostriker2010, kim2011, ostriker2011, kim2013, kim2015, kim2017}.

\begin{figure}[htb]
\centering     
\includegraphics[width=\linewidth]{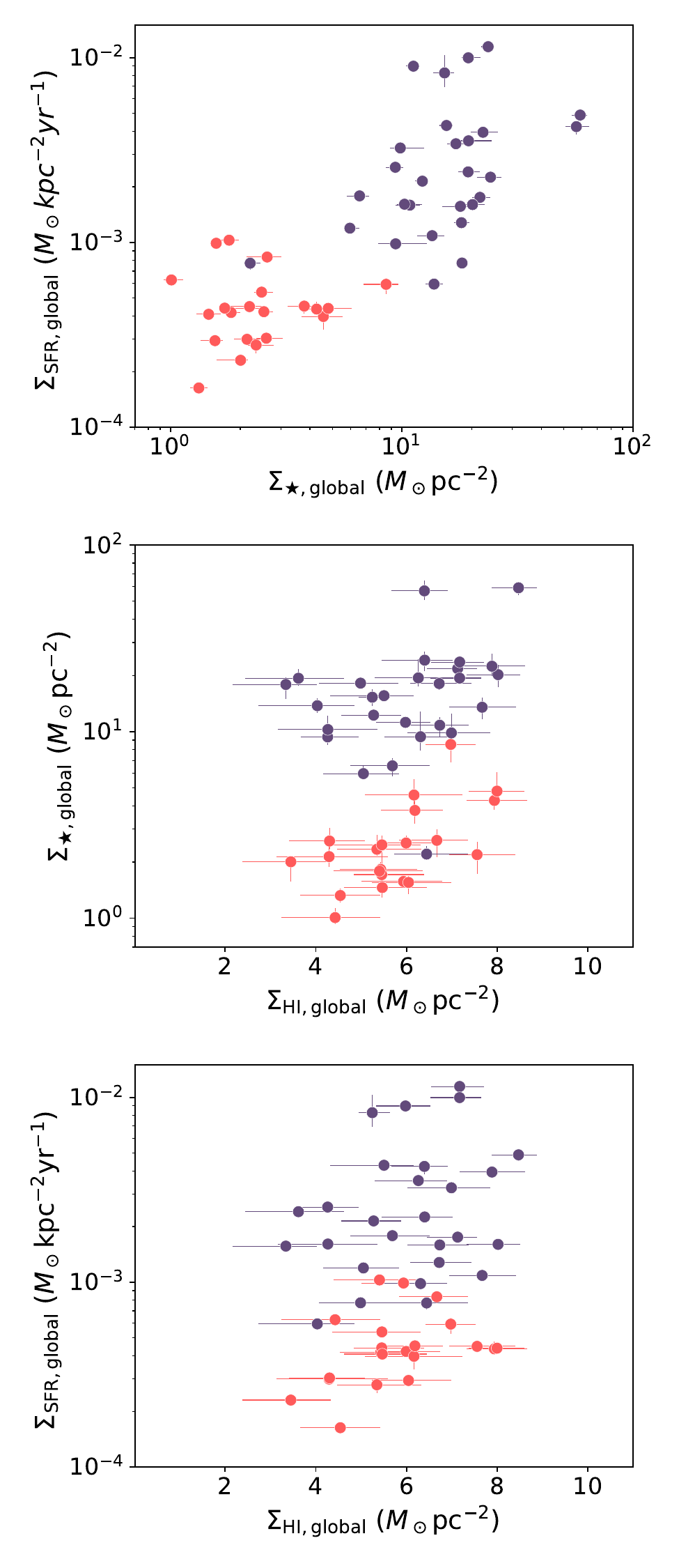}
\caption{ 
    Estimates of the SFR, stellar mass, and \HI{} surface densities in a globally
    averaged sense; i.e., the ratio between the integrated measurement and the
    isophotal size of the galaxy. UDGs are shown in red and NSA dwarfs in purple for
    all cases. In this figure, we exclude all NSA dwarfs with stellar masses
    exceeding $\rm M_\star = 10^9 M_\odot$. }
\label{f:specialornot}
\end{figure}

\section{Datasets}
Our dwarf sample consists of two main branches: a high surface brightness, ``normal'' dwarf sample
drawn from the \HI{} catalog of \cite{bradford2015} -- hereafter the
NSA sample, and an ultra-diffuse galaxy sample 
with known distances from the \hi{} catalog of \cite{janowiecki2019}.
The UDG sample is selected to have a maximum distance of $d=120$ Mpc; we
enforce the same limit on the NSA sample. We note that in this work we will
refer to the ``normal'' galaxy sample as the NSA sample, as these galaxies were
not specifically chosen to exclude low surface brightness galaxies. Rather, their relatively high surface brightnesses are a result of the observational selection which
lead to their inclusion in the NSA.

\subsection{Sample Overview}\label{s:methods:samples}
Here we detail the basic physical properties of the UDG and NSA samples. The sample consists of
58 galaxies at distances less than 120 Mpc; 
22 are UDGs from \cite{janowiecki2019} and 35 are NSA galaxies from
\cite{bradford2015}. All sources are required to have clean 
5-band HSC imaging, and all sources have measured \HI{} masses and redshifts due to their
selection method. The sample of \cite{janowiecki2019} is not explicitly chosen to be in the
field (in contrast to \citealt{leisman2017}), but they find that the properties of their \HI-selected
UDGs are largely unchanged by environment. We find that one UDG in our sample, AGC227965,
is quenched (SFR$<10^{-3}$ M$_\odot$ yr$^{-1}$) due to being a close satellite of MRK1324. We leave this galaxy in our
analysis as it still yields a significant \HI{} detection, but note that our discussion of the 
field UDG formation pathways
does not apply to this system.

In \autoref{f:gascomparison} we show
our sample galaxies in stellar mass versus \HI{} mass (left), SFR versus \HI{} mass (center), and
the star-forming main sequence (SFMS, right). These results
are from a customized SED fitting method that we describe in
detail in \autoref{s:methods:sedfitting}, but can be thought of as
standard SED fitting results for the purposes of the figures
in this section.
In this figure and all subsequent figures, we show the NSA
sample in purple and the UDG sample in red. The 
NSA sample reaches higher stellar masses and lower
\HI{} masses than does the 
UDG sample; when noted, we compare only NSA galaxies with
stellar masses within the observed range of the UDG sample.
The results of the \cite{leroy2008} sample of nearby
galaxies and SDSS spectroscopic value-added catalog \citep{kauffmann2003, brinchmann2004, salim2007}
are shown in beige and grey, respectively. We see that the UDGs have high \HI{} masses for their 
stellar masses and that they have low SFRs for their \HI{} masses -- conversely, the UDGs lie 
directly on the SFMS. 

A clear question arises from this initial analysis: are the \HI-selected
UDGs truly drawn from a more \HI-rich population than the NSA dwarfs, or are the higher \HI{} masses
simply an effect of observational selection?
To test this, we consider what our expected distribution of UDG \HI{} fractions, $\rm f_{HI} = M_{HI}/(M_\star+M_{HI})$,
would be
given the 21cm flux limit of the UDG sample if the $\rm f_{HI}$ distribution of
the UDGs were the same as the NSA sample. 
To put it another way, we ask the following question: if the UDG 
population had the same distribution of \HI{} fractions as the NSA dwarf sample, what
is the distribution of $\rm f_{HI}$ that we would expect to see from an
\HI-selected sample?

We show the results of this test in \autoref{f:fhi_detection}. The top panel
shows the observed 21 cm line fluxes in solid points (as always, purple for the
NSA sample and red for UDGs), while the unfilled red points show the expected
21 cm flux for each UDG in the sample if the \HI{} fraction were drawn from the
NSA dwarf distribution over $\rm f_{HI}$. The errorbars in this panel are generated
via bootstrap resampling from the NSA distribution. 
We estimate the detection limit of the UDG sample in 
two ways: first, from the stated completeness limit of the ALFALFA survey
for sources with $1.65 < \log_{10}(W_{50}/[\rm km\ s^{-1}]) < 1.7$ 
\citep[][black dashed line]{haynes2011}, and second, from the lowest 21 cm flux
of the UDG sample (red solid line). These limits are in good agreement with one
another, and we adopt the minimum 21 cm line flux of the UDG sample as our detection
limit for this exercise. The bottom panel of \autoref{f:fhi_detection} shows the
distribution of observed \HI{} fractions for the NSA (solid purple histogram) and
UDG (unfilled red histogram) samples, as well as the UDG sample that we would 
expect to have detected if the UDG distribution over $\rm f_{HI}$ were the same 
as that of the NSA sample (hatched dashed red histogram). Indeed, we find that 
we would expect to detect a significant number of 
$\rm f_{HI} \lesssim 0.75$ UDGs, should they exist.
Rather, the marked \HI{} richness of the UDG sample cannot simply be
attributed to observational limits.

We extend this positioning of \HI-rich UDGs as unusual star-forming environments by considering
a globally averaged stellar mass, \HI{}, and SFR surface density in \autoref{f:specialornot}. In this figure, we compute $\rm \Sigma_{\rm HI, global}$ using the total \HI{} mass and the
total area (as defined by the global elliptical HSC-g aperture, which we will describe
\autoref{s:methods:sedfitting}). 
One can immediately see that, as expected from their definition, the UDGs have significantly
lower SFR and stellar mass surface densities than the NSA sample. Intriguingly, however,
the UDGs do \textit{not} also show signs of lower \HI{} surface densities. Indeed, their \HI{} surface
densities span the same range as their spectroscopic counterparts. 

The main thrust of this paper
will be to root out the physical culprits of this apparently lowered SFE(\HI). 
From globally averaged measurements alone, it is unclear whether the 
low average star formation rate surface density is a result of 
a truly lower SFE(\HI), or whether the spatial distribution of star formation
in UDGs differs from normal dwarfs (e.g. a low globally averaged \sigsfr{}
could arise from an old stellar population that is significantly more
extended than the star-forming regions).
In order to distinguish between these two scenarios,
we must build an SED fitting method to constrain local (500 pc scale) SFR and stellar mass surface
densities (\autoref{s:methods:sedfitting}) -- we will return to this narrative of star formation
efficiency in \autoref{s:results}.


\begin{figure}[htb]
\centering
\includegraphics[width=\linewidth]{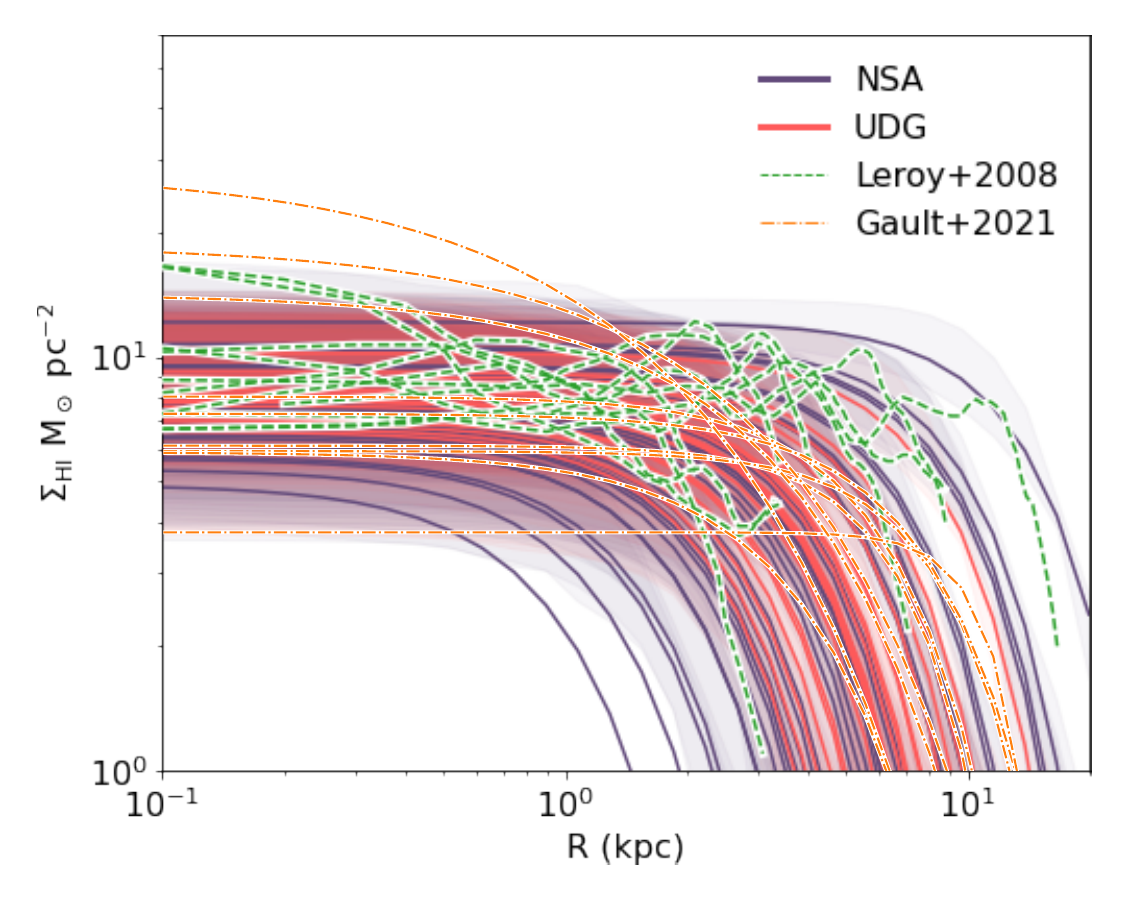}
\caption{
    A comparison between our estimates of the \HI{} surface density radial profiles for the
    UDG (red solid curves) and NSA (purple solid curves) samples against spatially resolved
    \HI{} measurements of dwarfs from the literature. We show the direct radial profiles 
    of \cite{leroy2008} in green, and \sersic{} fits inferred
    for a sample of \HI-rich UDGs from the
    \HI{} spatial distribution metrics provided by \cite{gault2021} in orange. We find that
    our radial profile estimates are in good agreement with these direct literature
    measurements.
}
\label{f:HI_profiles}
\end{figure}

\subsection{UV-Optical Imaging}

\textit{HSC-SSP.} The HSC-SSP imaging boasts wide spatial coverage, 
high sensitivity (a point source depth of $i_{\rm HSC} \sim 26$ in its shallowest ``Wide'' layer), and high
spatial resolution (the worst-seeing band, $g_{\rm HSC}$, has an average seeing of 0.77\arcsec{}).

For this work, we use the internal data release \hscdr{}, which covers 670 deg$^2$ in 5 optical bands. 
The HSC-SSP data products include two options for
the background subtraction -- a local subtraction suited
for small, high redshift sources, and a global subtraction
suited for nearby galaxies. We
use the 
\textsf{coadd/bg} versions of the co-added HSC images, which uses the ``global'' background
subtraction of the S18A and PDR2 data releases (for an in-depth discussion of the HSC global background
subtraction scheme, see Section 4.1 of \citealt{aihara2019} and
\citealt{li2021}). The data have been shown to reach
surface brightness limits of $i_{\rm HSC}\sim28.5$ mag arcsec$^{-2}$ 
for radially averaged 
measurements around a known target \citep{huang2018, kadofong2020c}. 

\galex{}. To complement the deep and high resolution optical imaging of the HSC
imaging, we draw upon archival \galex{} imaging to measure global UV photometry
of our galaxies. We use the package \textsf{astroquery} to query the 
Barbara A. Mikulski Archive for Space Telescopes (MAST) for \galex{} FUV and NUV
observations that cover the position of each galaxy. We adopt the average
\galex{} PSFs in this work; the FUV and NUV PSFs have full widths at half maximum
(FWHM) of 4.2'' and 4.9'', respectively\footnote{http://www.galex.caltech.edu/researcher/techdoc-ch5.html}.
We measure our \galex{} photometry on the background-subtracted intensity
maps provided by the \galex{} pipeline, and estimate uncertainties assuming
Poisson noise for the counts.

For targets with more than
one \galex{} archival image, we prefer the data product with a longer total
exposure time. Two NSA galaxies (NSA ID 5109 and 30738) do not have FUV 
imaging; all galaxies have NUV imaging. Of our sample with FUV imaging,
13 galaxies are drawn from the \galex{} All-Sky Imaging Survey 
(AIS, $m_{AB}=20.5$) and
35 from the Medium Imaging Survey \citep[MIS, $m_{AB} = 23.5$,][]{martin2005}.
13 are drawn from guest investigator (GI) programs; we
make an estimate the depth of the
GI programs by measuring the standard deviation of random apertures of apertures
the size of the average \galex{} FWHM. The GI imaging is somewhat deeper 
than the MIS imaging, with a mean $5\sigma$ depth of $m_{AB}=24.3$. 
Similarly, 6 galaxies in our sample have NUV AIS imaging, 40 have MIS imaging,
and 9 have GI imaging. We repeat the same procedure to estimate the depth
of the GI imaging and find a mean $5\sigma$ depth of $m_{AB}=23.4$. We note that
the depths estimated here are not used in the analysis, and are simply meant
to contextualize the \galex{} imaging sources for the reader.

\subsection{\hi{} Measurements}\label{s:methods:hi}
All of the galaxies in the present work have integrated \HI{} measurements from
\citet[][UDGs]{janowiecki2019} or \citet[][NSA galaxies]{bradford2015}. However,
resolved \HI{} measurements exist only for a small subset of the UDGs identified
by \cite{janowiecki2019} -- 12 from \cite{gault2021}, 3 from \cite{leisman2017}. 

To estimate the radial profiles of the \HI{} in our UDG and NSA
dwarf samples, we leverage the empirical result that UDGs and
``normal'' dwarfs are similar in their \HI{} structure -- that is,
UDGs lie on the \HI{} mass-size relation and have typical average
\HI{} surface densities \citep{gault2021}. We use this result
to estimate \sersic{} profile parameters for our sample via
a parameter estimation from the spatially resolved measurements
of \cite{hunter2021} (who report \sersic{} parameter fits for their sample), as
well as the spatially resolved measurements of the UDGs in
\cite{gault2021}\footnote{here we solve for the \sersic{}  parameters of their galaxies using the given measures of \HI{}  structure}.
We perform a least squares fit to the
\sersic{} index, $n$, central surface density, $\Sigma_{0,\rm HI}$
and effective radius, $\rm R_{eff, HI}$, of the literature samples
as a function of \HI{} mass; only $\rm R_{eff, HI}$ shows a 
significant correlation with \HI{} mass (i.e. the \HI{} mass-size
relation).
This allows us to estimate \sersic{} parameters
for our galaxies as a function of their \HI{} mass. 
As these structural measures are not well-constrained,
we will
account for the uncertainties induced by our estimated
\HI{} profiles
to demonstrate to the reader that these estimates are 
indeed
reasonable.

First, although the \sersic{} profile has three free parameters, our knowledge
of the total \HI{} mass allows us to determine the value of one of these parameters
given estimates for the other two. We therefore choose to fix the \HI{} central
surface density via the analytic solution for the total \HI{} mass of a 1D 
\sersic{} profile, i.e.:
\begin{equation}
\begin{split}
    {\rm M}_{\rm HI} &= 2\pi \int_0^\infty \Sigma_{\rm HI}(R) RdR\\
    &= 2\pi \Sigma_0 n ( (\frac{1}{h})^{1/n})^{-2n} \Gamma(2n). \\
\end{split}
\end{equation}
\rrr{We find that the $\Sigma_0$
values derived from a
fit to the literature data and the $\Sigma_0$
values derived using the total \HI{} mass are
in good agreement, with the latter method
resulting in $\Sigma_0$ values that are
on average $\sim 8\%$ higher than the former.}

Second, to account for the substantial uncertainty induced by our
lack of spatially resolved \HI{}, we include an uncertainty in the parameter
estimation of the \sersic{} profiles equivalent to the standard deviation of the
literature parameter values for reference (literature) galaxies within 
0.5 dex of that of the target (NSA/UDG) galaxy in total \HI{} mass. If no galaxies
are sufficiently close in \HI{} mass, we increase the tolerance from 0.5 dex to
up to 0.9 dex, but note that because the reference sample of \cite{hunter2021} is
systematically lower mass than our sample, this only affects the most massive
($\log_{10} \rm M_{HI}>10$) galaxies in our sample. We then estimate the
uncertainty on the radial \HI{} surface density profile via Monte Carlo 
resampling assuming that the uncertainties 
are well-describe by normal distributions.

The resulting \HI{} surface density profiles are shown in \autoref{f:HI_profiles};
as above, red curves indicate UDGs and purple curves
indicate NSA galaxies. The shaded regions span the 16\thh{} to 84\thh{} percentiles
of the Monte Carlo simulation result for each galaxy. The directly measured \HI{} 
profiles of \cite{leroy2008} are shown in green, while the \sersic{} profiles
inferred from the structural measurements of \cite{gault2021} are shown in orange.
We find that our inferred profiles are in good agreement with the shape, 
normalization, and scatter of the directly measured profiles, with the exception of
one UDG (AGC238764, which is not a part of our sample) that has an anomalously
high \sersic{} index ($n=1.98$; almost all dwarfs in this mass range are characterized
with sub-exponential \sersic{} indices, see \citealt{hunter2021}).
\rrr{There is, to our knowledge, only one
\HI-rich UDG that has spatially resolved
\HI{} data at a high enough resolution to 
probe the clumpiness and substructure of the
\HI{} \citep{mancerapina2022}. These data, 
which have a physical beam size of $\sim 2.9$
kpc, show significant \HI{} substructure
that is not captured in our simple radial
profile estimation approach. However,
the deviations from the radial profile that these azimuthal
variations cause are comparable to the uncertainty we have already
attributed to our radial profiles (see right panel of
Figure 1 of \citealt{mancerapina2022}). We thus find it unlikely
that we are overestimating the precision of the \HI{} profiles, 
though we of course note that obtaining high resolution
\HI{} maps would dramatically improve the precision
and accuracy of our $\Sigma_{\rm HI}$ estimates.}

\begin{figure*}[htb]
\centering     
\includegraphics[width=\linewidth]{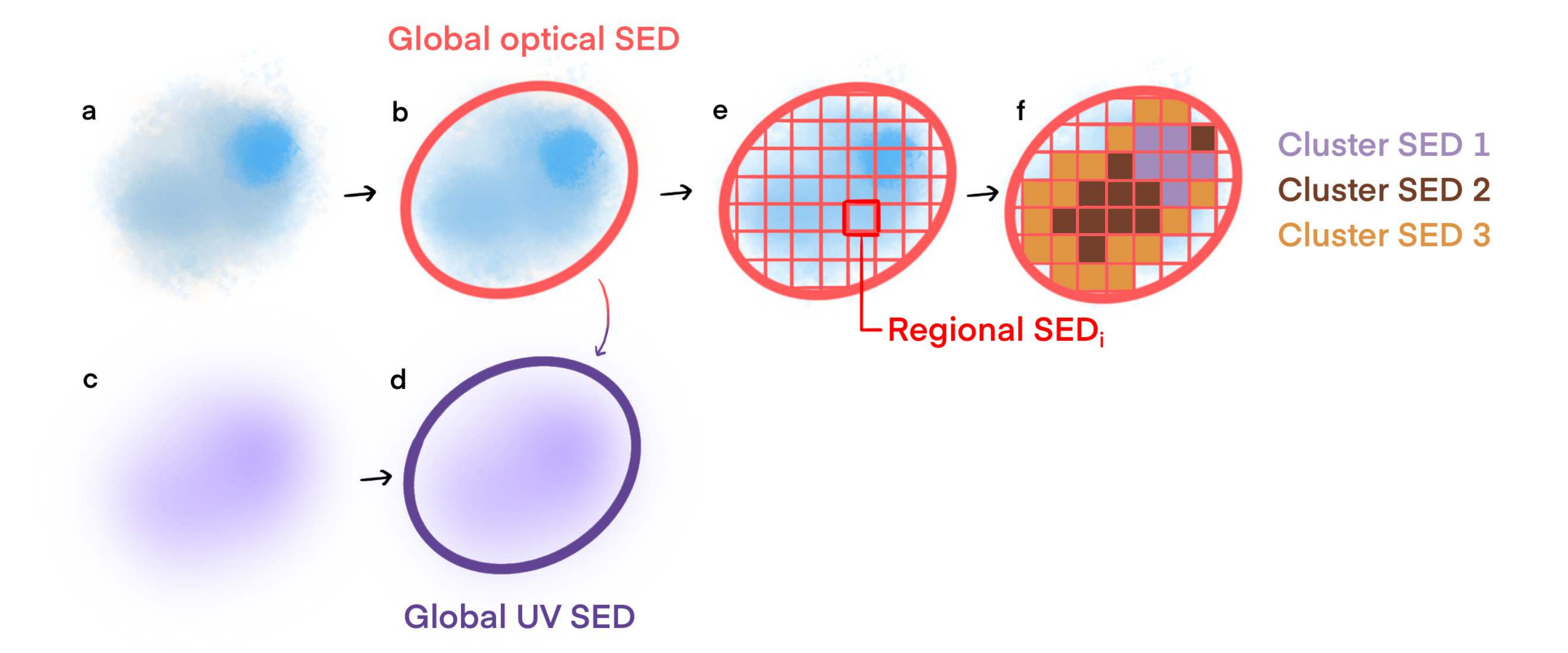}
\caption{ 
    A schematic layout of the fitting technique used in this work. At left,
    we represent the high-resolution optical data and
    low-resolution UV data (panels \textit{a} and \textit{c}, respectively). Panels
    \textit{b} and \textit{d} represent the global fit between the optical and 
    UV data; in this step, the apertures are matched between the optical and UV data.
    Panel \textit{e} shows the spatial division of the optical data into 
    regions in which regional SEDs are measured. Panel \textit{f} shows the clustering of
    these spatial regions into three representative clusters; these cluster SEDs will
    be fit jointly with the global UV photometry (panel \textit{d}).
    }
\label{f:schematic}
\end{figure*}

\begin{figure*}[htb]
\centering     
\includegraphics[width=\linewidth]{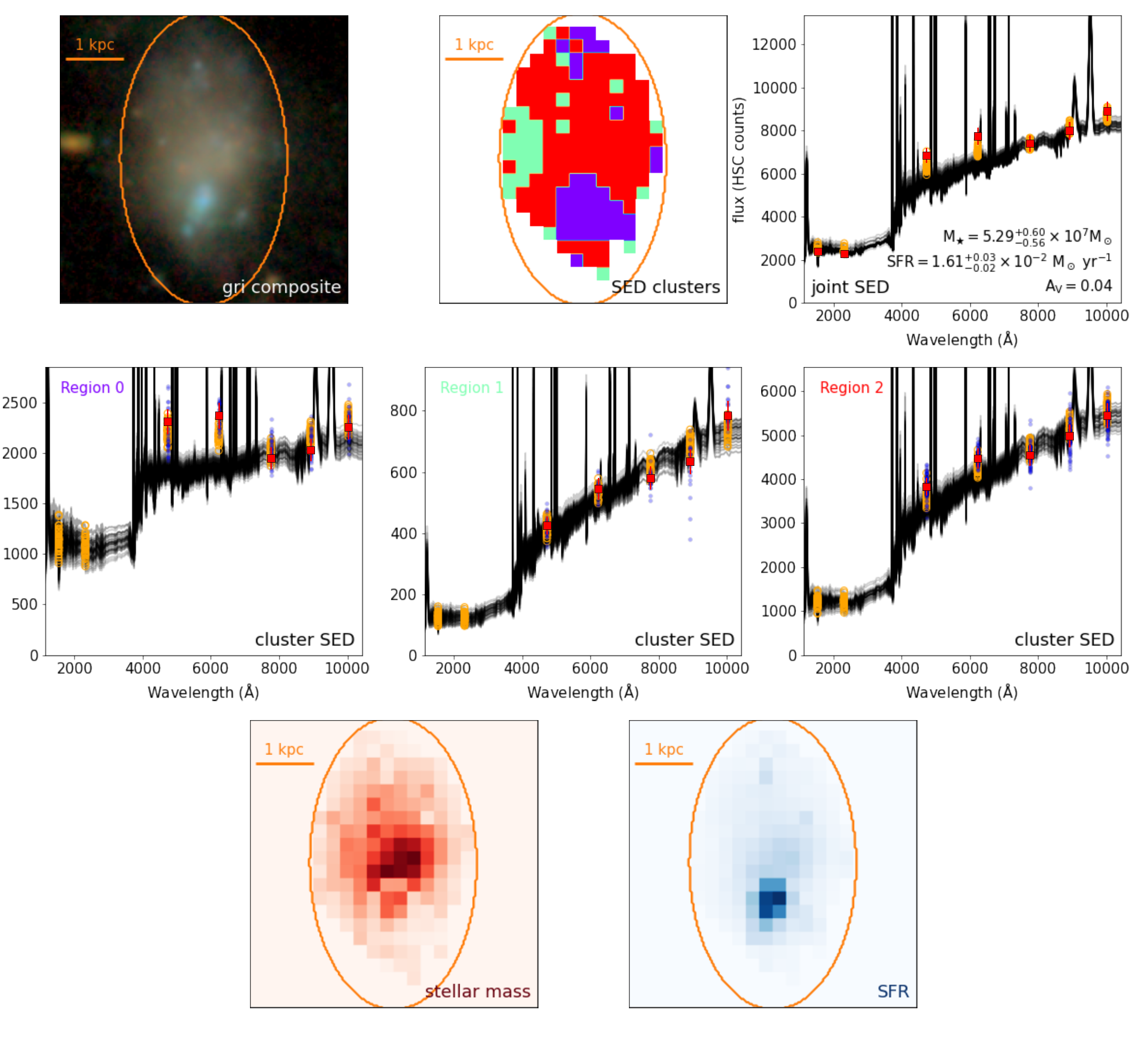}
\caption{ 
    An example of the fitting process for one of the NSA galaxies (NSA ID 5449). 
    \textit{Top row:} from 
    left, the HSC $gri$-composite image, the SED cluster map (analogous to panel \textit{f} 
    of \autoref{f:schematic}), and the joint UV-optical SED fit. The joint model SED is
    the result of fitting the spatially resolved (cluster) optical photometry and global UV 
    photometry. The black curves show the model spectra, while the orange points show the
    model photometry for each filter. The red points show the global UV-optical photometry.
    The model fitting and figure fluxes are computed as HSC counts.
    \textit{Middle row:} optical-only results for the individual cluster SEDs. The 
    color of the title text corresponds to the color of the region in the top middle panel;
    the format is equivalent to the joint SED fit (top right), with the addition of the
    SEDs measured in individual spatial regions \rrr{(i.e. regional SEDs)} as blue points. 
    \rrr{Here, the red points show the cluster SEDs (i.e. the sum of the regional SEDs) rather
    than the global photometry}.
    \textit{Bottom row:} the resultant stellar mass (left) and SFR (right) maps. 
    }
\label{f:bradford_example}
\end{figure*}

\section{SED Fitting}\label{s:methods:sedfitting}
The key technical challenge of this work is inferring the star formation and stellar mass properties of the
dwarf sample at sub-galactic scales. We accomplish this by modeling the high spatial resolution \hsc{} optical
data in conjunction with the UV \galex{} measurements to constrain star formation activity in a consistent
way with the UV data.

The novelty of our approach is to 
leverage both the high spatial resolution of \hsc{} and the expanded 
wavelength coverage of \galex{} to infer local star formation surface density 
properties of our dwarfs via a 
joint fit to localized optical photometry and global UV photometry. 
{}
Here, we detail the fitting process employed in this work. 

\subsection{Stellar Population Models}
We use the Flexible Stellar Population Synthesis (FSPS) models \citep{fsps_ref0, fsps_ref1}
with a fixed stellar and gas-phase metallicity of $Z=0.004$ 
along with a \cite{kroupa2001} 
initial mass function
to generate our synthetic spectra. 
\rrr{
We choose a single value for the model metallicity as the mass-metallicity relation is not well-understood
for these dwarf samples. Direct ($T_e$) measurements of nearby dwarfs \citep{lee2006, berg2012, jimmy2015}
can show 
significant offsets relative to each other and to metallicity estimates from SED fitting to more distant
dwarf samples
\citep{bellstedt2021}. We thus choose to adopt a typical value following the 
literature compilation of \citep{bellstedt2021} for a galaxy of $M_\star\sim 10^8 M_\odot$. We also 
check the impact of adopting solar metallicity models in \autoref{s:appendix:validation}, and find that our
SED fitting results are shifted by less than $0.1$ dex on average.
}
The library 
includes nebular emission, which may contribute significantly to the broadband flux of the dwarf galaxies in our sample. We adopt an exponentially declining star formation history 
\begin{equation}
    {\rm SFR} (t) \propto e^{-(t-t_0)/\tau};
\end{equation}\label{e:expdeclining}
We choose this relatively rigid star formation history (two free
parameters, the onset of star formation $t_0$ and the timescale of decay $\tau$) in order to achieve reasonable computation times when the number of regions to be fit is high.
Such a rigid choice of the SFH functional form has the potential
to bias SFH recovery, and so we note that the 
main goal of our analysis is to diagnose the relative change in star formation
surface density and star formation efficiency between normal, high surface brightness dwarfs and 
ultra diffuse galaxies, not to provide comprehensive stellar population synthesis
properties for the UDGs. We will return to this point more
comprehensively in \autoref{s:validation} when we consider 
validation tests for this method.
{}

The composite stellar population (CSP) 
model library spans ages between 2 Myr and 12 Gyr and timescales ($\tau$) between 0.3 Gyr and 10 Gyr. Each axis consists of 50 
logarithmically spaced values.
We pre-compute both the UV-optical spectra and the fluxes through the \galex{} and \hsc{} bands for each model in the library. During fitting, we linearly interpolate between pre-computed library fluxes to obtain the 
predicted \galex{} and \hsc{} fluxes for arbitrary choices of $t_0$ and $\tau$. Because we use pre-computed
library fluxes at a fixed distance of 10 Mpc, this induces a small k-correction error in the galaxy sample. However,
the most distant galaxies in our sample are at 120 Mpc, and so the k-correction error induced
is negligible.

\subsection{Defining Global, Regional, and Cluster SEDs}
The first step to executing a joint fit between the optical and UV photometry is to 
establish the apertures in which we measure the optical and UV photometry. 
To obtain consistent optical photometry, we PSF match the \hsc{} imaging with the worst 
seeing.
To do so, we use the Fourier transform-based implementation of \textsf{photutils} with a 
cosine bell window function ($\alpha=3$) to compute matching kernels between the other optical bands and
the band to which we will match the PSF.
We model the PSFs measured by the HSC pipeline (which produces
PSF postage stamps with spatial extents of about 7\arcsec{}) as two-dimensional Moffat functions. We choose
to simplify the PSF model to Moffat functions in order to prevent the emergence of imaging artifacts during
the matching process caused by high frequency features in the original pipeline-produced PSFs.
We additionally correct for galactic extinction in all UV and optical photometry presented in
this work using the measurements of \cite{schlaflyfinkbeiner2011} via the 
associated online service \citep{IRSAdust}.\footnote{Galactic 
extinction measurements were accessed via https://irsa.ipac.caltech.edu/applications/DUST/}

We then define the global aperture of the galaxy using \textsf{sep}, a python implementation of the segmentation
algorithm used in SExtractor \citep{bertin1996}. 
In all cases, we use the \hsc-g band as the detection band, with
a detection threshold of 5$\sigma$ over the variance image supplied by the 
\hsc{} pipeline. We
define the elliptical global aperture as
\begin{equation}
 c_{xx} (x-\bar x)^2  + c_{yy} (y-\bar y)^2 + c_{xy} (x-\bar x) (y-\bar y) = 3^2,
\end{equation}
where $\bar x$ and $\bar y$ are the light-weighted central coordinates 
of the galaxy and $c_{xx}$, $c_{yy}$, and $c_{xy}$ are the standard 
\textsf{SExtractor} ellipse parameters
\begin{equation}
\begin{split}
	c_{xx} &= \frac{\cos^2\theta}{a^2} + \frac{\sin^2\theta}{b^2}\\
	c_{yy} &= \frac{\sin^2\theta}{a^2} + \frac{\cos^2\theta}{b^2}\\
	c_{xy} &= 2 \cos\theta \sin\theta(\frac{1}{a^2} - \frac{1}{b^2}),\\
\end{split}
\end{equation}
where $a$ and $b$ are the semi-major and semi-minor axes, respectively, and $\theta$ is the 
position angle of the galaxy measured counterclockwise from the horizontal (in image coordinates).
This global aperture step is shown visually 
in step \textsf{b} of \autoref{f:schematic} -- the global aperture defined in \step{b} using
the optical data is then applied to the UV imaging to derive the global UV SED (\step{d}). 
We choose not to PSF match the optical data to the UV data to maintain the spatial resolution of
the optical data in order to hold the input images consistent between the global and
spatially resolved SED fitting. 
If we did perform PSF matching to the \galex{} imaging to compute the global
optical photometry, we have verified that the
changes would be negligible, with a mean fractional deviation in flux of 0.03 and a standard 
deviation of -0.02. 

Next, to define the regional boundaries for the optical photometry, we lay a grid of fixed size over the
cutout image, as shown in \step{e} of \autoref{f:schematic}. The box length of this grid is
set to be no less than twice the FWHM of the PSF-matched images, and is increased iteratively
until the median SNR of the SEDs measured in the regions is at least 3. To remove regions where
a reliable SED cannot be measured at the required spatial scales, we remove the bottom 5\% 
faintest regions. We also mask regions with $(g-i)>0.8$ and
generate additional masks for the UDGs, whose diffuse structure and
large on-sky sizes make 
standard deblending routines unreliable.
We verify that this pruning does not affect the total 
inferred mass and star formation rate in \autoref{s:validation}.

Because the size of the regions is small compared to the size of the galaxy, it is 
unsurprising to note that many of the regional SEDs are identical to each other to within their
photometric errors.
To decrease the number of parameters to fit, and to increase the SNR of the SEDs that
we fit, we group the regions (\step{e}) into clusters of similar SEDs (\step{f}). 
We group the regions into clusters using K-means clustering, initially with 3 clusters. 
If any cluster has a SNR$<10$, we combine the low-SNR cluster with the closest high-SNR 
cluster (here defining closeness by the $\ell^2$ norm\footnote{Where the 
$\ell^2$-norm is the difference between the SED flux and reference flux $\sqrt{\sum_k^K |f_k - f_{\rm ref}|^2}$ over 
the $K$ optical bands}  over the optical SEDs). Next, we must
verify that the cluster SEDs adequately represent their input regional SEDs. We use
the metric $z^j_{b,i} = |f_{b,i} - \bar f^j_b|/\sqrt{\sigma_{f_{b,i}}^2 + \sigma_{\bar f^j_b}^2}$, where
$f_{b,i}$ is the flux in band $b$ measured in region $i$, while $\bar f^j_b$ is the flux in
band $b$ measured in cluster $j$. $\sigma_{f_{b,i}}^2$ and $\sigma_{\bar f^j_b}^2$ refer
to the uncertainty on the respective flux measurement.
We impose the condition $z_{b,i}<2$ for all bands $b \in \{\rm g,r,i,z,y\}$ and all regions $i$ assigned
to cluster $j$ for each cluster in the galaxy. If the clustering result fails this verification,
we rerun the clustering process with an additional cluster at the K-means step. We allow
up to five clusters, at which point the clustering assignment is accepted regardless of
verification status. In practice, only one galaxy (the UDG 
AGC223246) is fit with the maximum five clusters.

\begin{figure}[htb]
\centering     
\includegraphics[width=\linewidth]{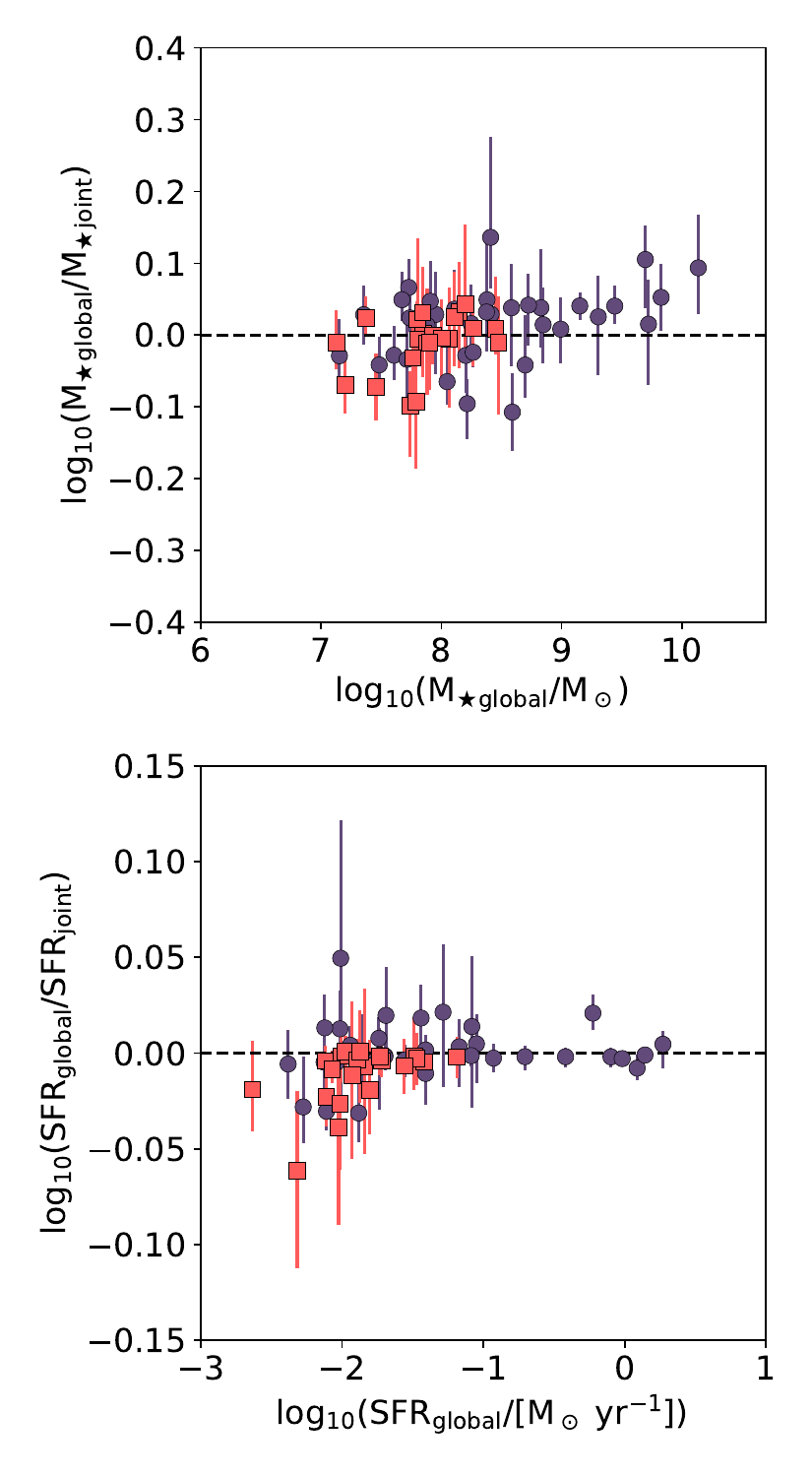}
\caption{ 
    \textit{Top:} a comparison between the best-fit stellar masses of the
    NSA (purple) and UDG (red) samples using the \rrr{difference between results derived
    from the spatially-resolved and} global photometry (y-axis) and
    the \rrr{global} optical photometry (x-axis). 
    \textit{Bottom:} the same as above, but for SFR.
    }
\label{f:internalconsistency}
\end{figure}

\begin{figure*}[htb]
\centering     
\includegraphics[width=\linewidth]{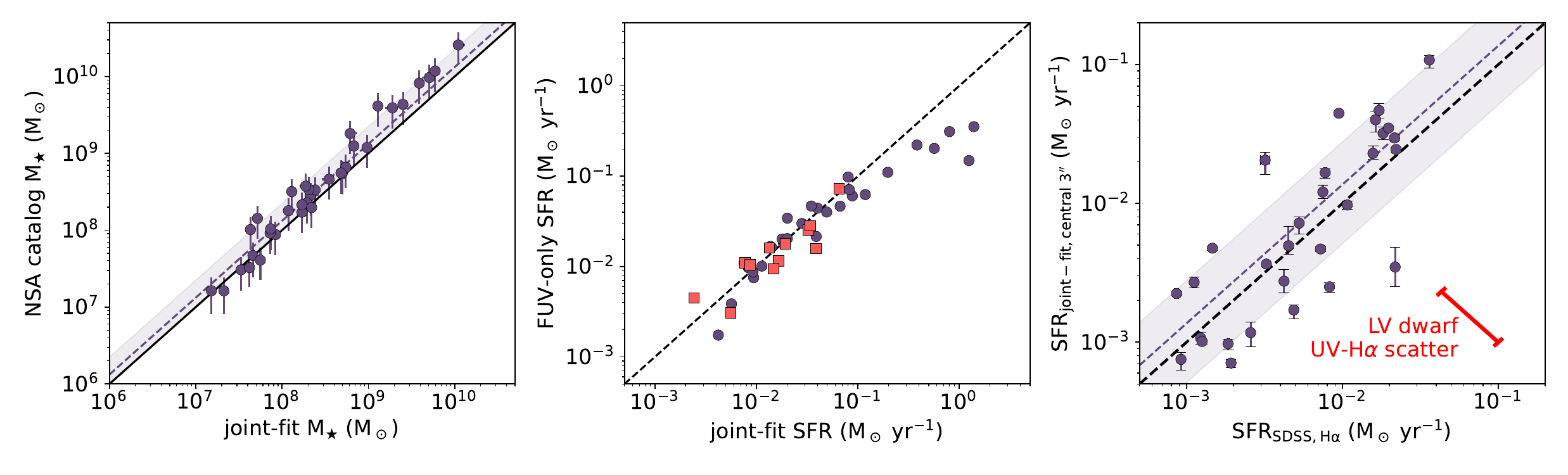}
\caption{ 
    \textit{Left:}
    a comparison between the stellar masses of the galaxies in this work (x-axis)
    and their catalog NSA stellar masses. The dashed line and shaded regions show the 
    median, and range between the 16$^{\rm th}$ and 84$^{\rm th}$ percentiles of the
    deviation $\log_{10}(\rm M_{\star}/M_\odot) - \log_{10}(\rm M_{\star,NSA}/M_\odot)$.
    \textit{Middle:} the same, but for SED fitting-derived SFRs and 
    FUV-only SFRs as computed from the catalog \galex{} measurements  
    using the SFR(FUV) prescription of \cite{kennicutt1998}.
    \textit{Right:} a comparison between SFRs derived from the 
    H$\alpha$ fluxes measured from SDSS 3'' fiber fibers (x-axis) and the 
    total SFR within the central 3'' of our SED fits (y-axis, red). 
    The dashed black line denotes a 1:1 relation. 
    As in the left panel, the dashed purple line and 
    shaded purple regions show the median, and range between the 16\thh{}
    and 84\thh{} percentiles. SFRs derived from H$\alpha$ measure star 
    formation over a shorter timescale ($\sim 10$ Myr) than our SED fits
    (100 Myr), and some physical scatter is expected as a result; 
    we show the standard deviation (in logarithmic space) of the 
    difference between H$\alpha$-derived and FUV-derived SFRs for a set of
    Local Volume dwarfs from \cite{karachentsev2021} 
    as a visual aid to the reader in gauging the scatter.
    } 
\label{f:compare2catalog}
\end{figure*}

\subsection{SED Fitting}
With our PSF-matched photometry and apertures in hand, we are now ready to begin the SED fitting. For
all of our fitting, we use the Markov Chain Monte Carlo implementation of \textsf{emcee}
\citep{foremanmackey2013}. We adopt a flat prior over 
positive values for the amplitude of each cluster SED, 
a flat prior between 0 and 13.6 Gyr for
$\tau$ (see \autoref{e:expdeclining}), and a log-normal prior over $A_V$ distributed
as $\log_{10}(A_V) \sim \mathcal{N}(0,0.1)$. In our joint fitting step, where multiple amplitudes
and $\tau$ values are being inferred, the prior is the same for each individual parameter. 

We first perform a global SED fit using both the optical and UV photometry to constrain the reddening, \av{},
and to provide an estimate of the total stellar mass and star formation rate. We adopt a Gaussian likelihood,
defined as:
\begin{equation}
\log \mathcal{L} \propto -\frac{1}{2} \sum_b^M  \frac{( f_b - \hat f_b )^2 }{\sigma_{f_b}^2} + \log(\sigma_{f_b}^2)
\end{equation}
where $f_b$ refers to the flux of the $b^{\rm th}$ band out of $M$ bands ($M=7$ for our 
fiducial case), $\hat f_b$ is the
predicted model flux, and $\sigma_{f_b}$ is the uncertainty in the $b^{\rm th}$ band. We measure $\sigma_{f_b}$ directly from the variance images
of HSC-SSP added in quadrature to an extra uncertainty of $0.05f_b$.
In this step we run \textsf{emcee} for 1500 iterations with 32 walkers, 
discarding the first 400 steps. The only parameter from this fit that survives as a final
result is the reddening, which is used to the joint fit as a fixed parameter. We find no difference in the $A_V$ estimated for the 
UDGs and the NSA galaxies at $M_\star < 10^9 M_\odot$: the
50\thh{} (16\thh{} and 84\thh{}) percentile of the $A_V$ distribution
is $A_V=0.17$ (0.04, 0.37) for the UDGs and $A_V=0.18$ (0.05, 0.38) for
the NSA dwarfs at $M_\star < 10^9 M_\odot$. Unsurprisingly, the
NSA galaxies at higher stellar mass have systematically higher
estimated reddening -- $A_V= 0.67$ (0.55, 0.99).

Next, we compute the flux in the $N$ SED clusters defined as above. At this stage,
the model consists of a set of $N$ models across $M$ bands. The likelihood is computed as follows:

\begin{equation}\label{e:jointlikelihood}
\begin{split}
\log \mathcal{L} \propto& -\frac{1}{2} \sum_{i \in UV} \frac{( \sum_j^N f_{i,j} - \sum_j^N \hat f_{i,j} )^2 }{\sum_j^N \sigma_{f_{i,j}}^2} + \log(\sum_j^N \sigma_{f_{i,j}}^2)\\
& + -\frac{1}{2} \sum_j^N \sum_{i\in Opt}  \frac{( f_{i,j} - \hat f_{i,j} )^2 }{\sigma_{f_{i,j}}^2} + \log(\sigma_{f_{i,j}}^2),
\end{split}
\end{equation}
where $i$ refers to the bandpass index and $j$ refers to the cluster index. In other words, we retain a 
Gaussian likelihood wherein the UV bands (FUV and NUV) are compared against the sum of the regional
models, whereas the optical bands ($Opt$ in \autoref{e:jointlikelihood}; g,r,i,z, and y) are compared against the appropriate regional models. In this step, 
the walkers are initialized with the positions drawn from the posterior of
the global SED fit; the amplitude of each cluster SED is additionally 
multiplied by the mean ratio between the flux in the cluster SED and the 
global SED over the optical bands.
In this step, we run \textsf{emcee} for 20000 iterations with at least 32 or $2N+1$ walkers,
discarding the first 5000 steps.

From this process we estimate the total stellar mass and star formation rate for each cluster
SED. We then convert these to stellar mass and star formation rate surface densities assigned
to each regional SED by multiplying by the ratio of the flux in the regional SED and the flux
in its corresponding cluster SED, and dividing by the physical area subtended by the region.

We show an example of the fitting procedure for the galaxy NSA 5449 in 
\autoref{f:bradford_example}. We also show an abbreviated gallery of 
fitting results in \autoref{s:appendix:gallery} for a set of galaxies that span
the morphologies seen in the NSA and UDG samples.

\begin{figure*}[htb]
\centering     
\includegraphics[width=\linewidth]{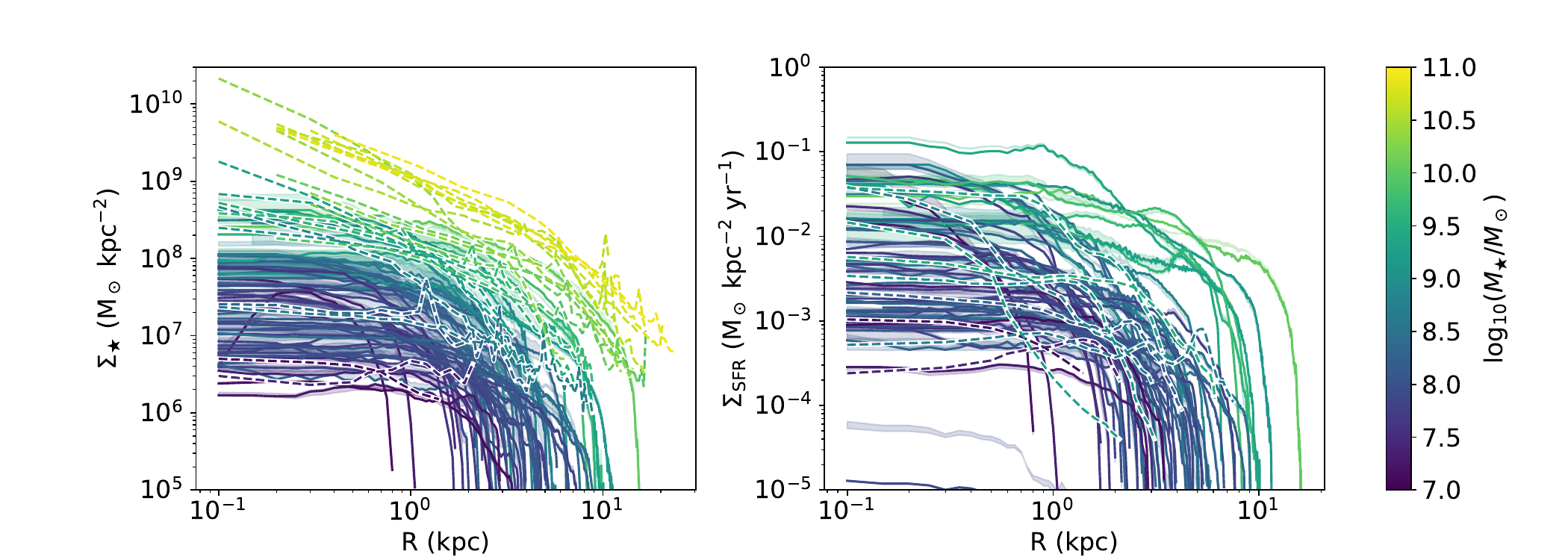}
\caption{ 
    A comparison between radial profiles in stellar mass surface density (left) and
    star formation rate surface density (right)
    from our SED fitting method (solid curves) as compared to direct measurements from 
    \cite[][dashed curves]{leroy2008}. The color of the curves show the integrated
    stellar mass of the galaxy in all cases. We find that there is good agreement between
    our radial profiles and the directly measured profiles in both amplitude and 
    extent.
    }
\label{f:leroyradialcomparison}
\end{figure*} 

\section{Method Validation}\label{s:validation}
In order to apply the framework constructed in the previous section to our galaxy samples, we
must first take the necessary step of validating that our method is able to accurately deliver
the quantities of interest to this study -- namely, maps of the star formation and stellar mass
surface densities down to the spatial resolution of HSC. 

An equally important step is to 
lay out the quantities that we are \textit{not} attempting to constrain with this modeling. 
The age, stellar metallicity, and detailed (flexible) star formation history of the galaxies are
all parameters that may be estimated from SED fitting.
However, it has been shown that the exponentially declining tau model is an imperfect model 
of star formation histories at large, both due to its overall rigidity and
that it, by definition, ties early and late star formation
\citep{simha2014, carnall2019}. 
Thus, we stress that our 
main goals, and the subjects of our verification tests, will be the estimation of the star formation
rate and stellar mass surface densities, and not a detailed interrogation of the star formation
history and enrichment of the galaxies in our sample.

\subsection{Intra-sample Comparisons}\label{s:validation:global}
We first consider an internal comparison between the star formation rate and stellar masses
measured from the global UV+optical SED fitting (\step{b} and \step{d} of \autoref{f:schematic})
and the integrated star formation rate and stellar masses obtained by summing the properties
inferred from the cluster SED fitting (\step{f}). In 
\autoref{f:internalconsistency}, we show the integrated
estimates of stellar mass (top) and star formation rate (bottom)
for the UDGs (red) and NSA galaxies (purple). In all cases,
we find good internal consistency between the initial SED fit
on the global photometry and the final model SED of the spatially
resolved method.

We additionally compare the stellar mass and SFR estimates 
obtained from our spatially resolved SED fitting method to
the catalog NSA stellar masses and FUV-only SFRs in 
\autoref{f:compare2catalog}. The FUV-only SFRs are computed 
from the catalog \galex{} measurements using the
prescription of \cite{kennicutt1998}, i.e.
\begin{equation}
    {\rm SFR}(L_{FUV}) = 1.4\times10^{-28} \frac{L_{FUV}}{\rm erg\ s^{-1}\ Hz}.
\end{equation}
Both the catalog stellar mass estimates and the FUV-only SFRs
are also in good agreement with the results of our spatially
resolved fit.

Although we lack truly spatially-resolved spectroscopic measurements for the 
galaxies in our sample, we do have access to one quasi-resolved measurement for
the NSA sample -- the SDSS 3'' fiber spectra. Because our galaxies exceed the
SDSS fiber size, the SDSS spectroscopic measurements do contain information 
about the spatial distribution of star formation within the galaxy, albeit in 
a very limited capacity.
In the right panel of \autoref{f:compare2catalog} we show a 
comparison between SFRs derived from the SDSS fiber spectroscopy of the 
NSA galaxies and SFRs we compute by integrating our SFR maps over SDSS-like
apertures. The SDSS spectroscopy we reference are 3'' diameter fiber spectra;
we use the NSA catalog line fluxes in this comparison. In particular, we use
the catalog H$\alpha$ flux to compute H$\alpha$-derived star formation rates
within the SDSS aperture following \cite{calzetti2013} with a \cite{kroupa2001}
IMF:
\begin{equation}
    {\rm SFR}(L_{H\alpha}) = 5.5\times10^{-42}\frac{L_{H\alpha}}{\rm erg\ s^{-1}}.
\end{equation}
Although the 100 Myr averaged SFRs we produce via our SED fitting are not
completely analogous to the H$\alpha$ SFRs derived from the SDSS spectra 
(which probe timescales closer to $\lesssim 10$ Myr), we expect that the two
should be well-correlated. Indeed, we find that the star formation rates we
compute from the central three arcseconds of our star formation rate maps are
in good agreement with the SDSS spectroscopic measurements, with a scatter
on the same order as the physical variation between UV and H$\alpha$ 
SFRs measured from Local Volume dwarfs \citep[][red line in right panel]{karachentsev2021}.

\subsection{Inter-sample Comparisons to the Literature}
We also find good agreement between our results
and literature measures of the integrated stellar mass,
SFR, and $\rm A_V$ for the two low surface brightness
galaxies presented in \cite{greco2018b}, though we do not include these
galaxies in this work because they lack \HI{} measurements.

Finally, we compare to spatially resolved measurements of stellar mass,
SFR, and \HI{} in nearby samples of dwarfs from the literature. In \autoref{f:leroyradialcomparison}, 
we show a comparison between radial profiles of stellar mass surface density (left) and
SFR surface density (right) as derived from our SED fitting method as compared to the directly
measured results of \cite{leroy2008}. Our results are shown as solid curves, the \cite{leroy2008}
profiles are shown as dashed curves; all curves are colored by total stellar mass. 
Encouragingly, we find good agreement in both the shape and normalization between our inferred radial
profiles and the directly measured results of \cite{leroy2008}, indicating that our method
is able to recover not just the total amount of stellar mass and SFR, but the distribution of
these quantities.

\begin{figure*}[htb]
\centering     
\includegraphics[width=\linewidth]{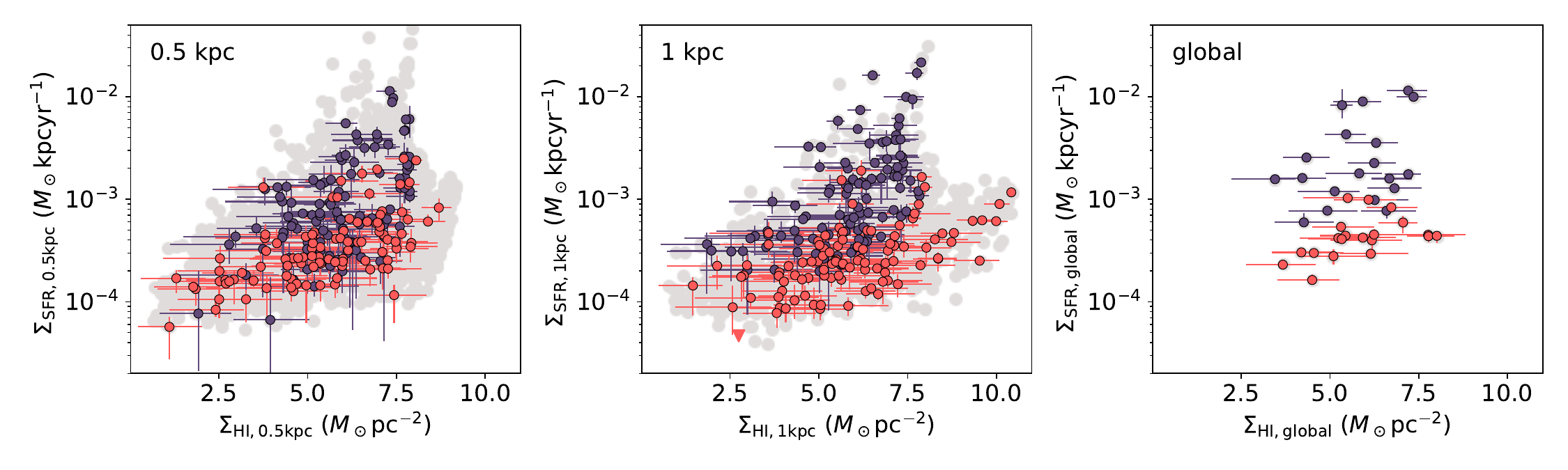}
\caption{ 
    A comparison between \HI{} surface density estimates and
    SFR surface density measurements averaged over 500 pc, 1 kpc, and
    global scales. 
    In all panels, we only show NSA galaxies that are within the stellar
    mass range of the UDG sample (i.e. $\rm M_\star<3\times10^8\ M_\odot$). 
    For visual clarify, we show a random subset of (at most) 
    70 points for each sample with individual errorbars and colored by 
    sample source; we show the full sample as the underlying grey scatter.
    Here, we demonstrate that the low SFE(\HI) that characterizes the
    globally averaged properties of the UDGs persists down to sub-kpc
    scales.
    }
\label{f:HIcomparison}
\end{figure*} 

\begin{figure*}[htb]
\centering     
\includegraphics[width=\linewidth]{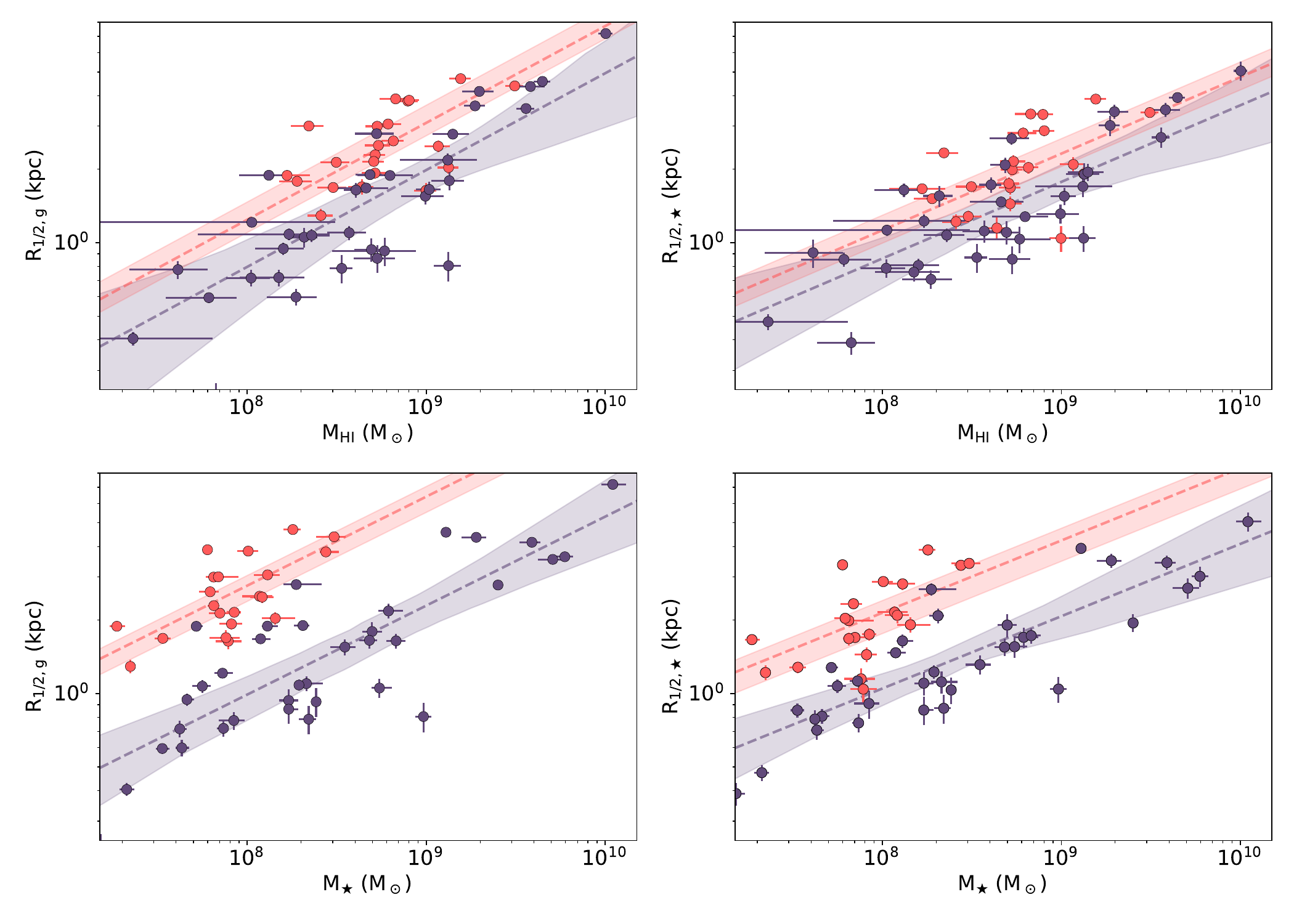}
\caption{ 
    A comparison between various measures of galaxy mass and galaxy size;
    the top row shows size as a function of \HI{} mass, while the bottom row shows size
    as a function of stellar mass. The left column shows size as measured by
    the HSC-$g$ band half-light radius, while the right column shows size as measured 
    by the half stellar mass radius. In each panel, the 
    purple dashed line shows the least squares power law fit to the NSA galaxies. 
    The red dashed line shows the least squares fit to the UDG sample when the index
    of the power law is held fixed to that of the NSA dwarf fit. The shaded regions 
    span the 16$^{\rm th}$ to 84$^{\rm th}$ percentiles as estimated from bootstrap
    resampling. As one expects, the UDG sample is large in size for their 
    stellar masses. However, their half-stellar mass radii are unremarkable for 
    their \HI{} masses.
    }
\label{f:HISMS}
\end{figure*}

\begin{figure}[htb]
\centering     
\includegraphics[width=\linewidth]{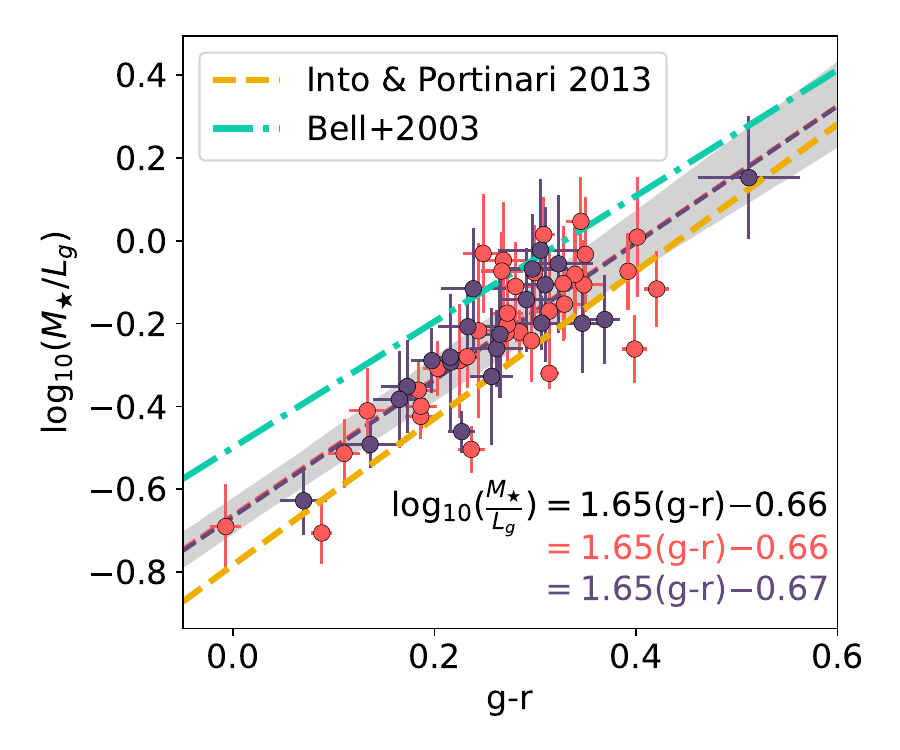}
\caption{ 
    Optical color versus stellar mass-to-light ratios in the HSC-$g$ band computed
    for our UDG (red) and NSA (purple) samples. We find no significant difference
    in the color-mass relation between the two samples, and additionally find the
    samples to be reasonably well-described by color-mass relations from the literature
    (\citealt{into2013} as the dashed line, \citealt{bell2003} as the dotted line).
    }
\label{f:lsbg_colormass}
\end{figure}

\begin{deluxetable*}{l|ccccccccc}
\tablewidth{0pt}
\tablecaption{Selected SED fitting results}
\tablehead{
\colhead{ID} &
\colhead{RA}  &
\colhead{Dec}  &
\colhead{Dist}  &
\colhead{$M_{\rm HI}$}  &
\colhead{$M_\star$}  &
\colhead{SFR}  &
\colhead{$\rm A_V$}  &
\colhead{$R_{\rm 1/2,\star}$}  &
\colhead{$R_{\rm 1/2,g}$}  \\
\colhead{} &
\colhead{[deg]} &
\colhead{[deg]} &
\colhead{[Mpc]} &
\colhead{ [$10^8\ \rm M_\star$] } &
\colhead{ [$10^8\ \rm M_\star$] } &
\colhead{ [$10^{-3}\ \rm M_\star\ yr^{-1}$] } &
\colhead { [mag] } &
\colhead { [kpc] } &
\colhead { [kpc] }
}
\startdata
NSA16938  &  140.433 &  4.374 &  85.81 &    $13\pm2.8$ &       $5_{-0.58}^{+0.66}$ &     $40_{-1.5}^{+1.9}$ &    $0.15_{-0.099}^{+0.12}$ &        $1.9\pm0.38$ &    $1.8\pm0.32$ \\
NSA11919  &  132.078 &  1.263 &  27.67 &  $1.3\pm0.42$ &    $1.3_{-0.092}^{+0.19}$ &  $5.7_{-0.24}^{+0.35}$ &      $0.15_{-0.1}^{+0.17}$ &        $1.7\pm0.19$ &     $1.9\pm0.1$ \\
NSA16993  &  143.969 &  4.017 &  81.05 &    $36\pm4.2$ &          $51_{-9}^{+7.8}$ &   $800_{-10}^{+8.5}$ &   $0.64_{-0.051}^{+0.059}$ &        $2.7\pm0.48$ &     $3.5\pm0.3$ \\
NSA13797  &  174.755 &  1.336 &  24.92 &  $1.9\pm0.58$ &  $0.43_{-0.034}^{+0.042}$ &     $35_{-1.1}^{+1.4}$ &  $0.079_{-0.035}^{+0.036}$ &       $0.71\pm0.12$ &   $0.6\pm0.093$ \\
AGC334353 &  345.863 &  1.704 &  71.40 &    $12\pm1.9$ &     $1.2_{-0.18}^{+0.19}$ &     $12_{-1.2}^{+1.1}$ &     $0.1_{-0.071}^{+0.11}$ &        $2.1\pm0.28$ &    $2.5\pm0.27$ \\
AGC334349 &  345.277 &  1.998 &  51.10 &  $5.2\pm0.83$ &    $0.82_{-0.11}^{+0.12}$ &   $10_{-0.78}^{+0.76}$ &  $0.087_{-0.064}^{+0.097}$ &         $1.4\pm0.2$ &    $1.9\pm0.19$ \\
AGC103435 &    3.838 &  1.075 &  28.50 &   $1.9\pm0.4$ &  $0.14_{-0.012}^{+0.015}$ &  $8.6_{-0.15}^{+0.19}$ &   $0.18_{-0.058}^{+0.059}$ &        $1.5\pm0.11$ &    $1.8\pm0.11$ \\
AGC198540 &  141.308 &  3.313 &  63.30 &    $13\pm1.8$ &     $1.4_{-0.23}^{+0.42}$ &    $13_{-0.8}^{+0.68}$ &    $0.14_{-0.092}^{+0.11}$ &        $1.9\pm0.26$ &      $2\pm0.24$ \\
\enddata
\tablecomments{\autoref{t:sedfits} is published \rrr{alongside this article} in its 
entirety in the machine-readable format.
A portion is shown here for guidance regarding its form and content.}
\end{deluxetable*}\label{t:sedfits}

\section{Results}\label{s:results}

With our methodology in hand, we may now return our focus to the physical question at hand. In
\autoref{s:methods:samples}, we demonstrated that although
the UDGs have low globally averaged stellar mass
and SFR surface densities, their \HI{} surface densities are comparable to those of the NSA galaxies. Now, we 
can consider the cause of the low SFE(\HI) of the UDGs
-- and in particular whether this inefficiency persists
to sub-kpc scales -- as
well as the relationship between UDG structure and \HI{}
content. For the interested reader, we include a table of selected SED 
fitting results for the sample in \autoref{t:sedfits}.

{}

\subsection{The Efficiency and Distribution of Star Formation in UDGs}\label{s:results:sfdist}
In the top row f \autoref{f:HIcomparison}, we show the relation between the 
estimated $\Sigma_{\rm HI}$ and $\Sigma_{\rm SFR}$ at 500 pc, 1 kpc, and global scales
from left to right (we note that, for reasons of visual continuity, the bottom right panel
of \autoref{f:HIcomparison} is identical in content to the bottom panel of \autoref{f:specialornot}). We compute 1 kpc and 500 pc average stellar mass
surface density and SFR surface density maps by placing a grid of length
1 kpc or 500 pc (at the distance of the galaxy) 
over the stellar mass and SFR maps produced via SED fitting. We do not
compute average quantities over length scales less than the regional SED
box size for each galaxy.
For the 1 kpc and 500 pc regions, the \HI{} surface density is estimated using the central position
of the region and the \sersic{} profile estimated for the galaxy as described in \autoref{s:methods:hi}. From this figure, the reader may immediately notice that the 
shift in $\Sigma_{\rm SFR}$ persists down to 500 pc;
there are no 
regions in the area-weighted census of the UDG distribution of SFR surface density that match the
vigorously star-forming SFR surface densities observed in the NSA dwarfs.

The persistence of the discrepancy of star formation activity outcomes as a function of \HI{} 
surface density indicates that it is not simply a matter of the distribution of star formation
that drives the apparently low star formation efficiency (as a function of atomic gas) in the 
UDG sample. Rather, it appears to be a generic property of star formation in our sample of UDGs. The question, then, of whether a more comprehensive understanding
of star formation can explain this low star formation efficiency will naturally
arise to the reader. This question is the central topic of \citetalias{papertwo}, 
in which 
we will examine the impact of the UDGs' peculiar stellar structure on their
star formation activity within the framework of pressure-regulated, feedback-modulated
star formation \citep{ostriker2010, kim2011, ostriker2011, kim2013, kim2015, kim2017}.

We note that the minimum region size over which we can measure \sigsfr{} and 
$\Sigma_{\star}$ is set by the distance of the galaxy and the angular scale over
which we measure our regional SEDs (panel e, \autoref{f:schematic}). In particular,
the galaxies for which we are able to measure 500 pc averages tend to be the
closest galaxies in our samples, and are thus skewed towards lower stellar masses.
However, we find that this effect does not impact the results presented here, as the
majority of galaxies affected are more massive than the UDGs in this sample. 
In particular, there are five galaxies within the stellar mass range of the UDGs
that are affected. Four are themselves UDGs (AGC 227965, AGC 322019, AGC 198543,
and AGC 238961), and one is a NSA dwarf (NSA ID 17750).

\subsection{The \HI{} Properties of the UDG Sample}\label{s:results:hiproperties}
In addition to their star formation efficiency, it is of interest to understand
the relationship between the stellar structure of these \HI-rich UDGs and their
integrated gas properites.
In \autoref{f:HISMS}, we show the
mass-size relation of the UDG and NSA samples for two definitions of ``mass'' and ``size'' (i.e.
four total combinations). The left column shows sizes measured as g-band half-light radii, while the right column shows sizes
measured as half-mass radii, where we refer to the radius that contains half
the \textit{stellar} mass of the galaxy as the half-mass radius. 
The half-light, half-mass, and half-SFR radii reported in this work are measured over the same area of the galaxy as our derived stellar mass and SFR maps (i.e. only in pixels where the stellar mass surface density is nonzero). We use the PSF-matched HSC images with the same machinery as the half-mass and half-SFR radius measurement. The half-light radii reported in this work are thus not completely analogous to effective radii measured using other methods (e.g. Sersic fitting). In particular, though there is no systematic offset between the two methods ($\bar \Delta R/R = -0.05$ for the full sample, and -0.06 (-0.04) for the NSA (UDG) subset), there is significant scatter ($\Sigma_{\Delta R/R} = 0.15$ for the full sample, and 0.13 (0.19) for the NSA (UDG) subset). We estimate 
the uncertainty in our half-mass and half-light radius measurements by adding 
in quadrature the physical size of the PSF FWHM and the standard deviation of the
radii due to the uncertainty in the stellar mass fits/photometry, respectively.
The top row shows \HI{} mass, while the bottom shows stellar mass.
In essence, one can think of the bottom left panel as the ``canonical'' mass-size relation (stellar
mass versus light-weighted size).

When one considers the size of the UDGs as a function of their integrated stellar mass, 
it is immediately apparent  
that they are systematically larger than the NSA sample in both 
their light-weighted and mass-weighted sizes. This is
unsurprising, of course, given that UDGs are selected for
their low surface brightnesses and large physical
sizes.
When one considers the UDGs as a function of
their integrated \HI{} mass, however,
the picture changes. In particular,
we see no significant difference in 
the relation between the \HI{} masses and half-mass (stellar) radii of the UDG and
NSA samples. There is, however, some evidence that the half-light radii of the
UDGs remain elevated with respect to the NSA sample at fixed \HI{} mass.


\subsection{The Color-Mass Relation of UDGs}\label{s:appendix}
Though not the main focus of this work, we are able to directly 
compare the color-mass relation of gas-rich UDGs and ``normal'' dwarfs
as a byproduct of the SPS fitting presented above.
Standard color-mass relations \citep[e.g.][]{bell2003,into2013}
have been assumed in the literature for 
color-based estimates of UDG stellar mass \citep[cf][]{mancerapina2019b, karunakaran2020, marleau2021}, but
it has not been directly shown that
these optical color relations are applicable to the UDG population. In particular, 
the canonical color-mass relations are typically calibrated for and tested on largely
massive galaxies; it is thus of practical importance for both the UDG community
and general dwarf galaxy community to test the applicability of these relations 
on the extreme pockets of the dwarf population (e.g. UDGs, Blue Compact Dwarfs, etc.).

In \autoref{f:lsbg_colormass} we show the global (i.e. light-weighted) g-band mass-to-light
relations of the dwarfs in this work (as usual,
red points indicate
the UDG sample while purple points indicate the spectroscopic dwarf sample). We calculate
stellar mass-to-light ratios by computing the $\rm g_{HSC}$ luminosity 
and stellar mass 
from the global optical SED of each galaxy (panel b in \autoref{f:schematic}). The $\rm g_{HSC} -r_{HSC}$ colors are measured
using the same SED used to derive the stellar mass-to-light ratio.

We find that there is no offset between the
\HI{}-selected UDG sample and the NSA dwarf sample, and that the 
samples are well-described by a relation that lies between the 
\cite{into2013} and \cite{bell2003} color-mass relations. For 
completeness, we list our ordinary least squares parameters
and standard errors here, where we define 
$\eta\equiv (M_\star/L_g) (L_\odot/M_\odot)$.
We avoid the standard choice of $\Upsilon$ as it is 
reserved for a parameter linking star formation and
midplane pressure in 
\citetalias{papertwo}.
\begin{equation}
    \begin{split}
        \log_{10}\eta_{\rm all} &= 1.59 \pm 0.14 (g-r) - 0.65 \pm 0.04\\
        \log_{10}\eta_{\rm UDG} &= 1.65 \pm 0.21 (g-r)- 0.66 \pm 0.06\\
        \log_{10}\eta_{\rm NSA} &= 1.55 \pm 0.20 (g-r)- 0.64 \pm 0.06,
    \end{split}
\end{equation}
the exact calibration of this relation is, of course, dependent
on both the photometric bands used and IMF assumptions.
We in particular note that
\cite{into2013} and \cite{bell2003} are calibrated for 
SDSS g and r bands, which are similar but not identical to 
HSC g and r bands). Regardless, we have demonstrated that the UDGs do
not lie systematically off of the color-mass relation of the NSA
dwarfs, and that both samples are in relatively good agreement
with previous relations from the literature.

\section{Discussion}
Taken altogether, we find that our sample of \HI-rich UDGs are characterized by
low star formation efficiencies 
(where SFE(\HI) $\equiv \Sigma_{\rm SFR}/\Sigma_{\rm HI}$)
down to sub-kpc scales, and that the UDGs are generally incapable of 
supporting the vigorous star formation seen in the ``normal'' NSA dwarfs. 
Intriguingly, although the UDGs are large in light-weighted size for their
total stellar mass, their stellar mass-weighted sizes are typical for their 
\HI{} masses. Finally, the stellar mass-to-light ratios of the UDGs appear 
consistent with those of the NSA dwarf sample.

\subsection{Implications for UDG Formation}\label{s:discussion:implications}
The results presented in this work suggest that the \HI-rich Ultra-Diffuse
Galaxies are unable to efficiently convert gas to stars or host vigorous star formation. Such low star formation rate surface densities indicate that UDGs
will continue to trail behind the stellar mass build-up of the general dwarf
population.
This finding, however, does not alone answer the question of \textit{why} certain galaxies fail to trigger vigorous star formation. 
Several mechanisms have been put forth in the literature to explain the
population of field UDGs. For each formation model, we will first
discuss the points in which our observations agree with the model predictions
before turning towards points of potential conflict.

\textit{Feedback-driven expansion.} We first consider the picture in which
field UDGs are formed via starburst-driven outflows which cause potential fluctuations
that in turn lead to net expansion of the stellar population 
\citep{elbadry2016, dicintio2017, chan2018}. We note to the reader that in this 
section we deal with results from both the Feedback In Realistic Environments (FIRE, \citealt{hopkins2014},
and FIRE-2, \citealt{hopkins2018})
and Numerical Investigation of a Hundred Astrophysical Objects (NIHAO, \citealt{wang2015}) simulations;
though both simulations point to feedback-driven expansion as the culprit of UDG 
formation, the feedback implementation between the
two simulations is not identical.

In the NIHAO simulation,
such UDGs are expected to retain systematically higher \HI{} fractions with 
respect to ``normal'' dwarfs, in good agreement with our results \citep{dicintio2017}.
However, two points of potential conflict emerge. First, the radial expansion of
the UDG stellar population should enforce a radial age gradient; that is, the 
older stars should experience more expansion due to having gone through a larger
number of inflow/outflow cycles. This picture suggest a relatively centrally 
concentrated distribution of star formation in UDGs, where the stars experience 
net expansion while the gas goes through cyclical phases of inflow and outflow. 
Such behavior is in qualitative conflict with the observation that UDG star formation
is both consistently lower than and more radially extended than star formation in
the NSA dwarfs (see lower panels of \autoref{f:leroyradialcomparison}). However,
to our knowledge a detailed study of the star formation extendedness and efficiency
has not been done for UDGs created by feedback-driven expansion. Further analysis
is necessary to confirm that feedback-driven expansion cannot reproduce this
effect (via, e.g. the substantial expansion of stars within 100 Myr of formation). 

However, another point of conflict exists between our observations and the 
feedback-driven expansion picture. Namely, in both FIRE and NIHAO the marked
expansion of the UDGs is driven by particularly prolonged and bursty star formation
histories \citep{dicintio2017, chan2018}. In this work, we have shown that the
lack of vigorous star formation in UDGs is linked to their low stellar mass
surface densities; that is, their low SFR surface densities are explainable by
their stellar and gaseous structure. Conversely, then, the structure of the UDGs
makes it difficult to envision a scenario in which they are able to power 
frequent and vigorous bursts of star formation at late times.
\rrr{This does not, of course, rule out UDG formation via 
star formation feedback if the star formation behavior of UDGs was markedly
different at high redshift than it is in nearby UDGs. Indeed, such a transition
has been predicted in models such as \cite{trujillogomez2022}. 
However, these new observations provide new constraints on models which predict 
star formation feedback to remain vigorous down to $z=0$.}

\textit{Early mergers.} Another proposed mechanism for field UDG formation is
the result of an early merger (more than 8 Gyr ago) that caused an increase in total
spin as well as a decrease in central SFR and a redistribution of star formation
to large radii \citep{wright2021}. Like our observed galaxies, the galaxies in the 
\textsc{ROMULUS25} simulation are \HI-rich for their stellar mass and lie on the
integrated star-forming main sequence. However, we do not find the lower central
specific star formation that \cite{wright2021} cite as a significant observable 
manifestation of this formation channel; indeed, the distribution of sSFR of 
the UDGs in our sample is not statistically different than that of the NSA sample
at any radius. 

\textit{High-spin halos.} Finally, we consider the proposal that UDGs are 
formed out of the high spin tail of the halo distribution 
\citep[see, e.g.][]{dalcanton1997, amorisco2016, rong2017, dicintio2019, liao2019}. This formation
scenario predicts that UDGs will host ordered \HI{} disks, in line with
resolved measurements from the literature \citep{mancerapina2019, mancerapina2020, gault2021}. 
\rrrtwo{There is some suggestion that in this formation scenario,}
UDGs \rrr{at $z=0$} 
should have generally lower \rrr{central volumetric} gas densities \rrr{and more radially
extended profiles}
than high surface brightness galaxies \rrrtwo{at fixed stellar mass} \citep{liao2019}.
This \rrrtwo{may be at} odds with \rrrtwo{observational findings,
given that the UDGs in our sample have similar \HI{} 
surface densities to the NSA dwarfs (see, e.g. \autoref{f:HIcomparison}, which excludes
NSA dwarfs with stellar masses larger than the range probed by the UDG sample)
and have been previously suggested to lie on the \HI{} mass-size
relation \citep{gault2021}. Together, these findings
imply similar volumetric gas densities between the UDGs and NSA dwarfs if there is no 
systematic difference in disk thickness between the UDGs and NSA dwarfs.}
\rrrtwo{However, the relationship between UDG and normal dwarf \HI{} disk thickness
remains observationally uncertain.} 
\rrrtwo{We thus underscore that we are not seeking to make a comprehensive comparison 
between the \HI{} structural properties
of observed UDGs and their simulated high-spin analogs.}

\rrrtwo{The upshot of the discussion
at hand, rather, is }that neither the early merger 
scenario nor the high-spin halo scenario are in conflict with the most salient
point of our analysis -- that once set on a path towards low-efficiency
star formation,
\rrrtwo{galaxies like our present-day UDGs} are likely to continue to fall behind the stellar mass build-up
of the general dwarf population. 

\subsection{The Stellar Structure of Dwarfs at fixed $M_{\rm HI}$}
In addition to a comparison with established theoretical models, it is 
informative to consider what fully empirical statement we may make about the
relationship between the \HI-rich UDGs and ``normal'' NSA dwarfs. 
In the preceding sections, we have demonstrated that the stellar half-mass
radii of the UDGs are typical for their \HI{} masses 
(see \autoref{f:HISMS}) -- that is, the UDGs
are not offset from the mass-size plane for a particular casting of
mass (\HI{} mass) and size (stellar half-mass radius). This is contrasted,
naturally, to the offset in the normal mass-size plane (stellar mass and half
light radii) by which UDGs are definitionally characterized.

A concordance in the \HI-mass-stellar-size relation would naturally arise
if the UDGs were forming stars normally with respect to the spatial distribution
of gas, but at decreased star forming efficiency. 
To illustrate this idea,
let us presume that the spatial distribution of \HI{} in UDGs is normal with 
respect to high surface brightness galaxies, as has been shown for a number of
the UDGs in our sample by \cite{gault2021}. 
As hinted to in the introduction, the driving mechanism behind the 
low star formation efficiency of the UDGs will be further explored in
\citetalias{papertwo}.

In this picture, a galaxy that will become a UDG 
will form stars over the same spatial distribution as a galaxy that will become 
a high surface brightness galaxy, but at lower $\Sigma_{\rm SFR}$ -- 
thus becoming a galaxy with a lower total stellar mass. The
concordance of the UDG and NSA samples in the \HI{} mass-stellar size plane is
then a natural consequence of the comparatively lower star formation 
efficiency of the UDGs. 
Thus, one mechanism by which to 
generate the ``normal'' sizes of the UDGs with respect to their \HI{} masses
is for
present-day UDGs to have assembled most of their stellar mass via the
low SFE(\HI) star formation that characterizes their behavior at $z=0$.

\section{Conclusions}
Due to their large sizes and low stellar mass surface densities, 
UDGs are both an extreme 
outcome of galaxy formation and an extreme environment for star formation.
\HI-selected UDGs are systematically more gas-rich than their ``normal'' dwarf analogs
(see \autoref{f:fhi_detection}), and there is evidence that some UDGs have assembled the
majority of their stars from extremely dense clumps \citep{danieli2021}. 
Understanding the way in which star formation proceeds in UDGs is thus of key 
interest to both the galaxy evolution and star formation communities.

In this work, we have presented a method that allows us to compute star formation rate
and stellar mass surface density maps over galaxies with known distances using 
spatially resolved optical data and global UV photometry. We use this method to explore
the star formation activity in a sample of nearby (d $<120$ Mpc)
\HI-detected ultra-diffuse galaxies dwarf galaxies from \cite{janowiecki2019}, along with
a NASA Sloan Atlas reference sample of ``normal'' dwarfs with \HI{} measurements 
from \cite{bradford2015}.
The samples in this work allow us to compare and contrast the star formation behavior of the UDGs
against the NSA dwarfs, providing
new clues to the evolutionary pathway of these \HI-rich UDGs as well as
new tests for star formation theory in extreme (low density)
environments that we will explore in \citetalias{papertwo}.

We find that the globally averaged SFR surface densities of the UDGs are
low compared to their globally averaged \HI{} surface densities (see \autoref{f:specialornot});
this offset could be explained either by a truly lower SFE(\HI) in the UDGs, or by a difference
in the relationship between the distribution of the UDGs' stellar mass and SFR. Our results indicate
that the SFE(\HI) of the UDGs remains lower than the NSA dwarfs down to 500 pc scales (see 
\autoref{f:HIcomparison} and \autoref{s:results:sfdist}) -- the UDGs are indeed less efficient
at converting atomic gas (to molecular gas) to stars. In the 
second paper of this series, \cite{papertwo}, we will consider the role of
the galactic structure of the UDGs on their star formation. Indeed, we find that
the diffuse structure of the UDGs can naturally explain their low star 
formation rate surface densities and efficiencies. When the structure of
the UDGs is considered (rather than the gas content alone), we find that their
star formation is quantitatively consistent with star formation in both 
the NSA sample and in more massive galaxies. As such, we direct the interested
reader to \citetalias{papertwo} of this paper series.

Returning to the results at immediate hand,
we also consider the relationship between the stellar size and \HI{} mass of the UDGs, as
compared to the NSA galaxies (\autoref{s:results:hiproperties}). We find that although the
UDGs have large stellar sizes (both light-weighted and mass-weighted) for their total stellar mass,
their stellar sizes are largely unremarkable as a function of their \HI{} mass. This 
superficially surprisingly result can be explained naturally when we consider the previous 
conclusions of this work: the UDGs have fairly normal \HI{} structure and properties, and so the
spatial distribution of their star formation is set by those normal gas properties.
However, because they form stars at a significantly lower SFE(\HI), they form fewer stars overall and
thus become large for their total stellar mass.

The low efficiency and low star formation rate surface densities of the 
UDGs indicate that the UDGs will continue to lag behind the general
dwarf population in their stellar mass build-up, and thus remain 
diffuse relative to ``normal'' dwarfs. However, this does not answer the 
question of \textit{why} present-day UDGs first entered the evolutionary 
path to become UDGs at $z=0$.
While this work alone cannot provide a 
definitive answer to this question, we may consider the implications of our results on 
contemporary ideas of UDG formation. The 
spatially extended star formation of the UDGs and apparent inability to power vigorous star 
formation (high SFR surface densities)
are in conflict with the idea of 
UDG formation via the feedback-driven radial expansion of stars
\rrr{for the picture in which feedback-driven expansion continues even when stellar
mass surface density is low}. 
Formation mechanisms which 
set certain halos down the path towards UDG formation at early times (e.g. the high spin halo
scenario or early merger scenario) can more naturally explain the low SFE(\HI) cycle seen in
the sample of UDGs presented in this work. 

A natural extension of this work will be to obtain a larger number of spatially resolved \HI{} 
measurements of the UDGs in this sample, with a particular focus on the most nearby galaxies
in the sample (such that we may maximize the physical spatial resolution). 
Higher resolution UV-optical data will also allow us to directly probe the physical scales
relevant to star formation and cluster populations in these galaxies. 
A further exploration into the star formation and ISM of these UDGs is crucial to
further our understanding of UDGs as an extreme environment for star formation, as well as our
understanding of how these extreme objects form.

\begin{acknowledgements}
The authors thank Eve Ostriker, Chang-Goo Kim, and Lachlan Lancaster
for insightful comments and discussion that have greatly improved this manuscript. 
\rrr{We also thank the anonymous referee for their review of this work, which improved the
quality of the manuscript.}
The research of 
EKF was supported by the Porter Ogden Jacobus Fellowship. JEG gratefully 
acknowledges support from NSF grant AST-2106730.

Based in part on data collected at the Subaru Telescope and retrieved from the HSC data archive system, which is operated by Subaru Telescope and Astronomy Data Center at National Astronomical Observatory of Japan.

The Hyper Suprime-Cam (HSC) collaboration includes the astronomical communities of Japan and Taiwan, and Princeton University. The HSC instrumentation and software were developed by the National Astronomical Observatory of Japan (NAOJ), the Kavli Institute for the Physics and Mathematics of the Universe (Kavli IPMU), the University of Tokyo, the High Energy Accelerator Research Organization (KEK), the Academia Sinica Institute for Astronomy and Astrophysics in Taiwan (ASIAA), and Princeton University. Funding was contributed by the FIRST program from Japanese Cabinet Office, the Ministry of Education, Culture, Sports, Science and Technology (MEXT), the Japan Society for the Promotion of Science (JSPS), Japan Science and Technology Agency (JST), the Toray Science Foundation, NAOJ, Kavli IPMU, KEK, ASIAA, and Princeton University. 

This paper makes use of software developed for the Large Synoptic Survey Telescope. We thank the LSST Project for making their code available as free software at  http://dm.lsst.org

The Pan-STARRS1 Surveys (PS1) have been made possible through contributions of the Institute for Astronomy, the University of Hawaii, the Pan-STARRS Project Office, the Max-Planck Society and its participating institutes, the Max Planck Institute for Astronomy, Heidelberg and the Max Planck Institute for Extraterrestrial Physics, Garching, The Johns Hopkins University, Durham University, the University of Edinburgh, Queen’s University Belfast, the Harvard-Smithsonian Center for Astrophysics, the Las Cumbres Observatory Global Telescope Network Incorporated, the National Central University of Taiwan, the Space Telescope Science Institute, the National Aeronautics and Space Administration under Grant No. NNX08AR22G issued through the Planetary Science Division of the NASA Science Mission Directorate, the National Science Foundation under Grant No. AST-1238877, the University of Maryland, and Eotvos Lorand University (ELTE) and the Los Alamos National Laboratory.

Some of the data presented in this paper were obtained from the Mikulski Archive for Space Telescopes (MAST). STScI is operated by the Association of Universities for Research in Astronomy, Inc., under NASA contract NAS5-26555. Support for MAST for non-HST data is provided by the NASA Office of Space Science via grant NNX13AC07G and by other grants and contracts. 

Based on observations made with the NASA Galaxy Evolution Explorer. GALEX is operated for NASA by the California Institute of Technology under NASA contract NAS5-98034. 
\rrr{The specific observations analyzed can be accessed via \dataset[https://doi.org/10.17909/h19r-ab55]{https://doi.org/10.17909/h19r-ab55}.}

This research has made use of the NASA/IPAC Infrared Science Archive, which is operated by the Jet Propulsion Laboratory, California Institute of Technology, under contract with the National Aeronautics and Space Administration. 

This research has made use of the VizieR catalogue access tool, CDS, Strasbourg, France. 

\software{Astropy \citep{astropy:2013, astropy:2018}, matplotlib \citep{Hunter:2007}, SciPy \citep{jones_scipy_2001}, the IPython package \citep{PER-GRA:2007}, NumPy \citep{van2011numpy}}, 
pandas \citep{McKinney_2010, McKinney_2011},
Astroquery \citep{astroquery}, extinction \citep{barbary2021}
\end{acknowledgements}

\bibliography{ksdwarfs.bib}

\begin{appendix}
\section{Fitting Gallery}\label{s:appendix:gallery}
Here we include an abbreviated gallery of fitting results for the NSA and UDG
samples. We choose three galaxies from each sample with their median star formation
rate surface density and median stellar mass surface density maps; these galaxies
are chosen to roughly span the morphologies seen in the samples.

\begin{figure*}[htb]
\centering     
\includegraphics[width=.95\linewidth]{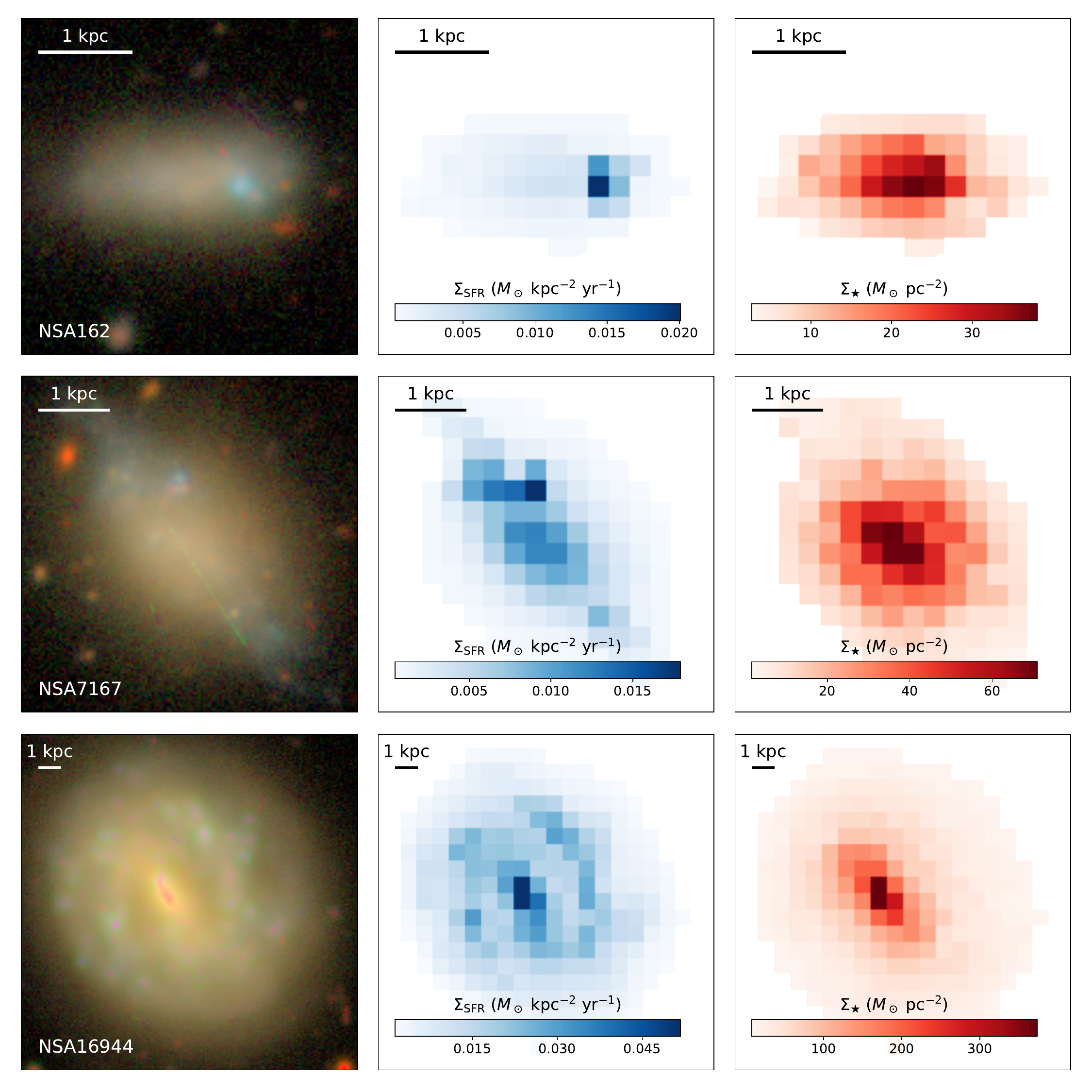}
\caption{ 
    A sample of three example galaxies from the NSA sample, as labeled. 
    From left, we show the HSC-SSP gri-composite RGB image, the 
    median star formation rate surface density map, and the
    median stellar mass surface density map. These galaxies are chosen to 
    span the range of morphologies present in the sample.
    } 
\label{f:NSA_gallery}
\end{figure*}

\begin{figure*}[htb]
\centering     
\includegraphics[width=.95\linewidth]{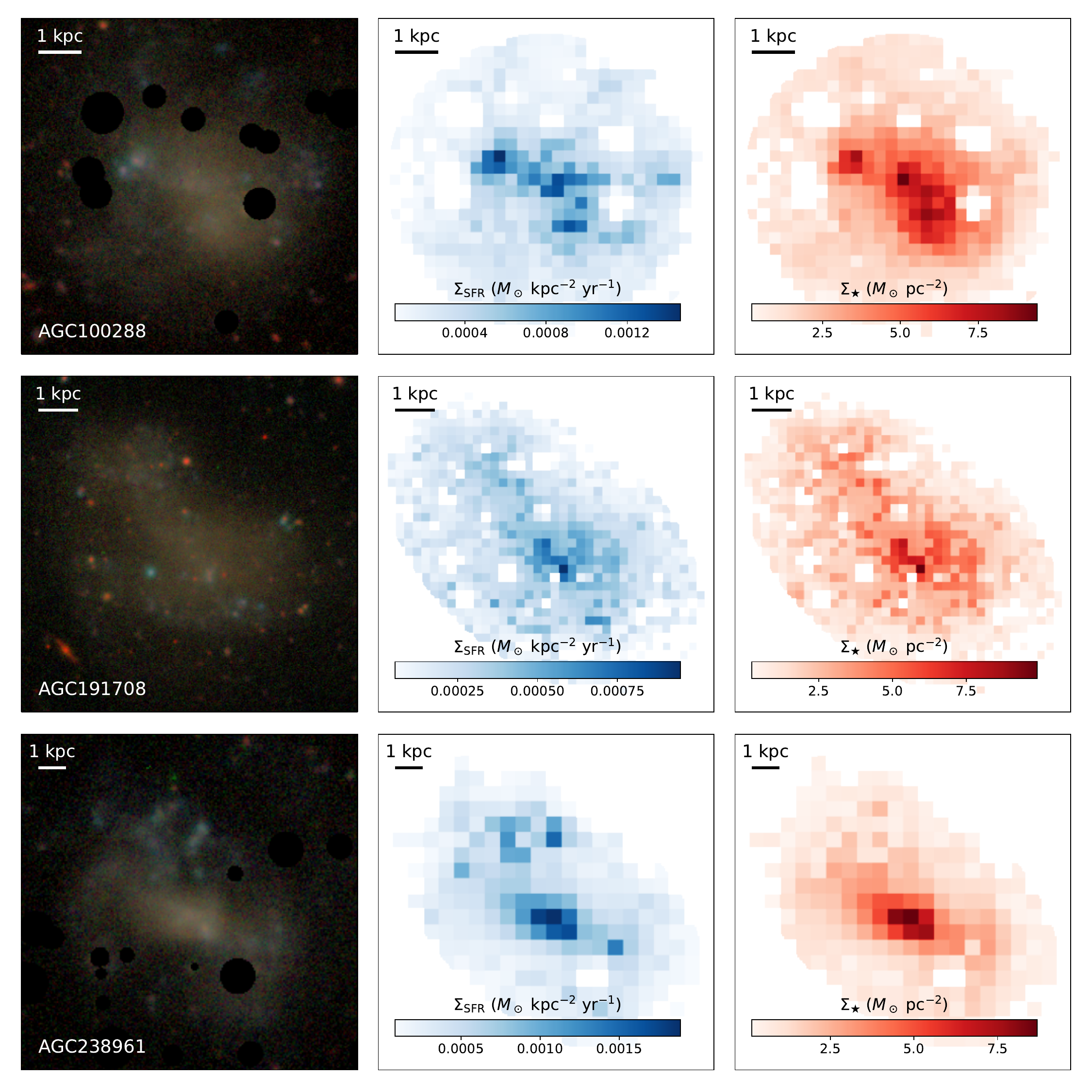}
\caption{ 
    A sample of three example galaxies from the UDG sample, as labeled. 
    From left, we show the HSC-SSP gri-composite RGB image, the 
    median star formation rate surface density map, and the
    median stellar mass surface density map. These galaxies are chosen to 
    span the range of morphologies present in the sample.
    } 
\label{f:AGC_gallery}
\end{figure*}

\begin{figure*}[htb]
\centering     
\includegraphics[width=.95\linewidth]{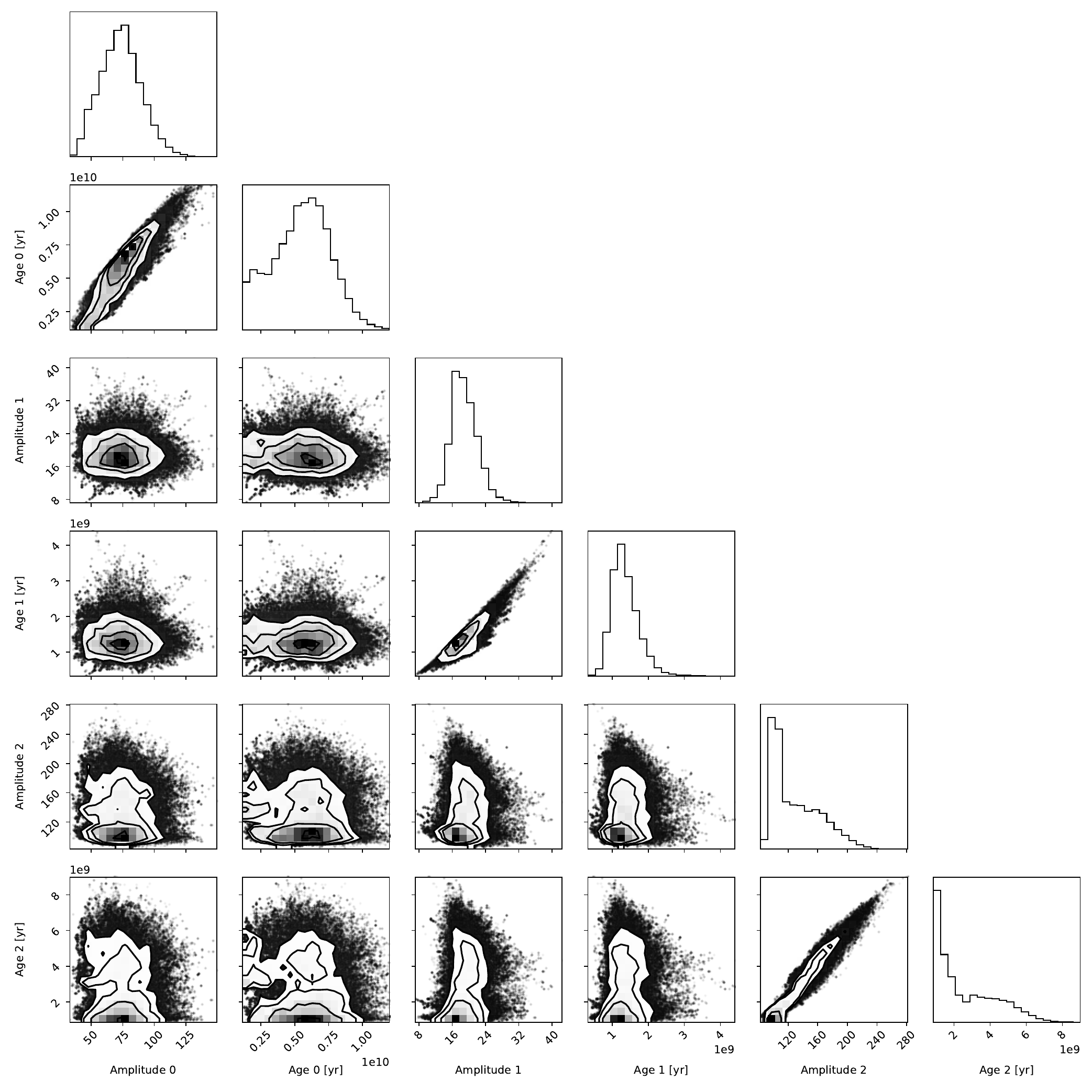}
\caption{ 
    \rrr{
    A cornerplot of the parameter inference for NSA7167. As expected,
    although the 1D distribution of the parameters are strongly 
    peaked and unimodal, there is some degeneracy in the tails of the distribution
    between the age and amplitude of each component. This can 
    be understood naturally by considering that as the age of a component
    increases, the mass-to-light ratio drops, meaning that 
    the amplitude of that component (i.e. the stellar mass) must increase
    to account for the same luminosity.
    }
    } 
\label{f:cornerplot_7167}
\end{figure*}

\begin{figure*}[htb]
\centering     
\includegraphics[width=.95\linewidth]{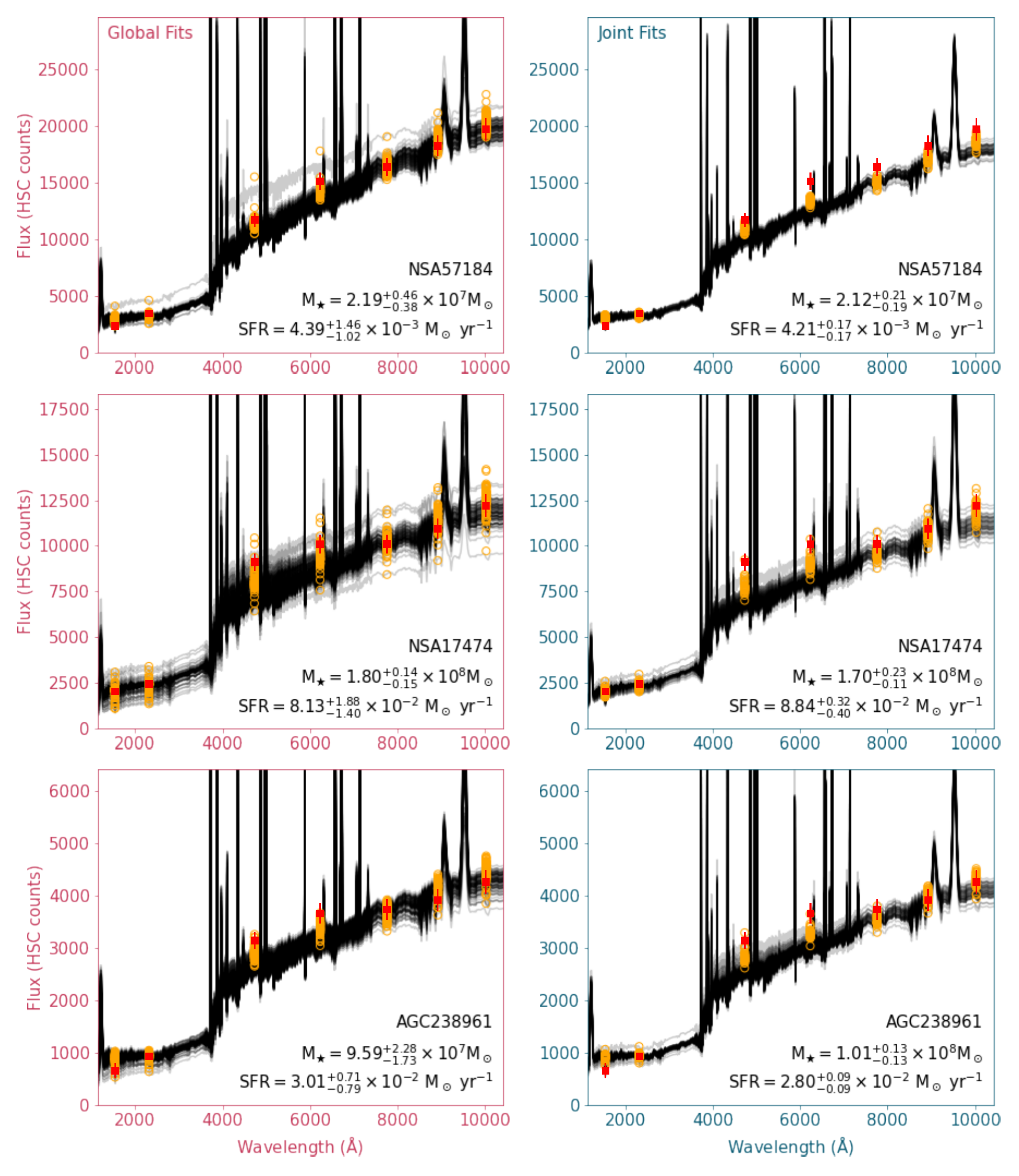}
\caption{ 
    \rrr{
    A comparison of the global SED fits (left column)
    and joint SED fits (right column)
    for three galaxies in our sample. In each panel, the
    red points show the observed fluxes in FUV, NUV, g$_{\rm HSC}$,
    r$_{\rm HSC}$, i$_{\rm HSC}$, z$_{\rm HSC}$, and y$_{\rm HSC}$.
    Each black curve and set of orange points show the spectrum
    and synthetic photometry, respectively, of one draw from the
    posterior. The text shows the median stellar mass and
    SFR estimates for each fit; uncertainties correspond to the
    16\thh{} and 84\thh{} percentiles of the posterior. 
    }
    } 
\label{f:fitcomparison}
\end{figure*}

\begin{figure*}[htb]
\centering     
\includegraphics[width=.95\linewidth]{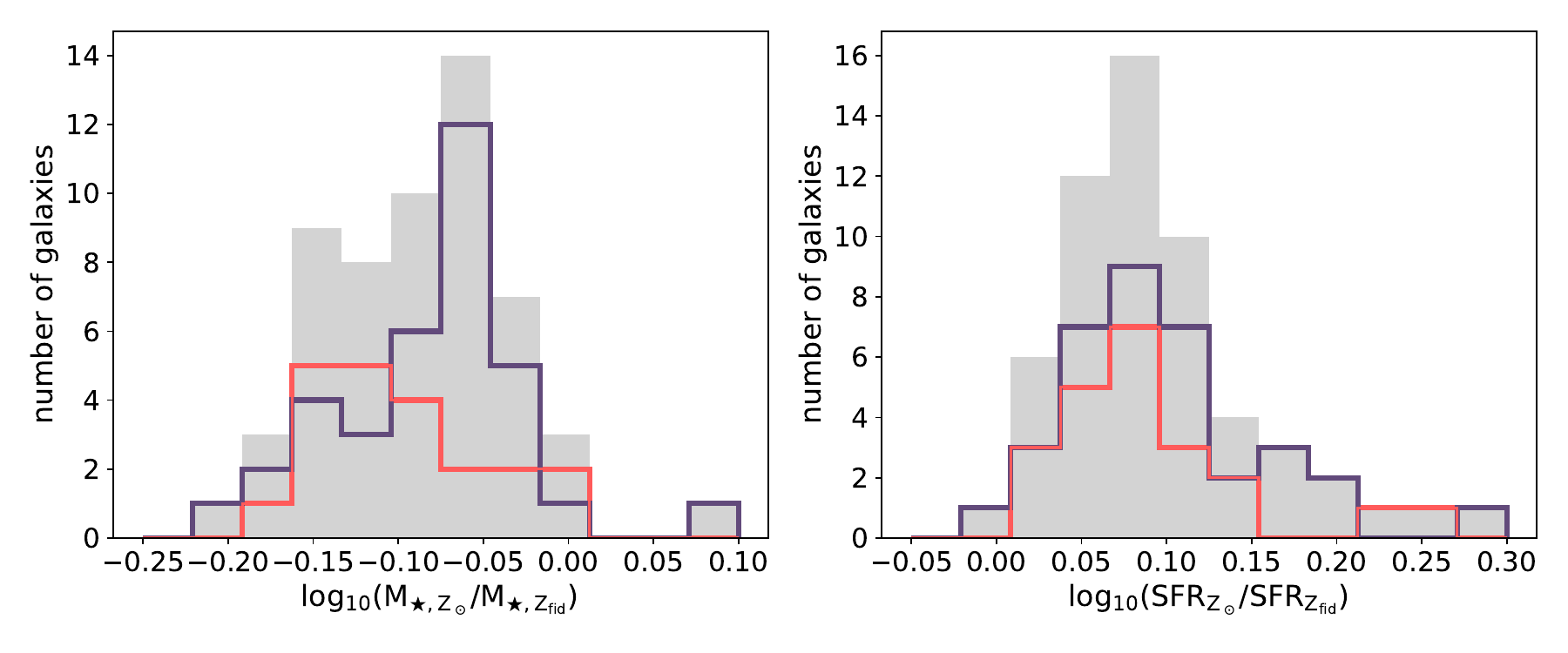}
\caption{ 
    \rrr{
    A comparison between the total stellar mass (left) and star formation rate (right)
    estimates derived from our spatially resolved SED fitting if we assume a metallicity of
    $Z=0.012$ (as opposed to our fiducial value of $Z=0.004$). In both panels, the grey filled
    histogram shows the distribution over all galaxies -- the unfilled red and purple histograms 
    show the distribution over the UDG and NSA samples, respectively.
    }
    } 
\label{f:cornerplot_7167}
\end{figure*}

\section{Additional Validation of the SED Fitting}\label{s:appendix:validation}
\rrr{
Due to the novelty of our joint-fitting approach, we describe three 
additional validation steps here. First, we show a typical
cornerplot for one of the jointly fitted SEDs (the example shows NSA7167) 
in \autoref{f:cornerplot_7167}. We find that the inferred parameters
are strongly peaked and unimodal, though there is a degeneracy between the
age and amplitude (stellar mass) inferred for each cluster SED. 
This is a physically
expected effect since an increase in the age leads to a drop in the
mass-to-light ratio of the stellar population, meaning that more 
stellar mass is necessary to achieve the same luminosity. 
We find no significant degeneracy between parameters inferred for
different cluster SEDs.
}

\rrr{
Next, we show in \autoref{f:fitcomparison} a comparison
between the SED fitting results using the global optical photometry
and those derived using the cluster SEDs.
We find no evidence for a systematic shift between the global
and joint fitting methods. Indeed, the total stellar mass
and star formation rate inferred via the joint fit method is 
in good agreement with those estimated via the global fit method,
as shown by the text in each panel.
}

\rrr{
Finally, as stated in the main text, we assume a fixed metallicity in the SED 
fitting presented in this paper. There is some evidence that low surface brightness galaxies
may have higher than average gas-phase metallicities \citep{greco2018b, sanchezalmeida2018}, 
though we caution that these results contain only three low surface brightness galaxies, and
that these low surface brightness galaxies are an optically-selected sample (as opposed to the 
\HI-selected UDGs in the sample discussed in this work). We find that assuming solar metallicity
models shifts the total star formation rate and stellar mass estimates by less than 0.1 dex compared
to the fiducial estimates. The stellar masses estimated using solar metallicity models are slightly 
lower (with a median difference between solar metallicity and fiducial values of -0.08 dex), 
and the star formation rates are slightly higher (with a median difference of 0.09 dex). This shift is
expected -- the higher metallicity models are redder at fixed age, meaning that the best-fit 
stellar populations will be on average slightly younger at fixed color. 
}

\section{\rrrtwo{\HI{} Profile Recovery}}
\rrrtwo{Although the subset of the \cite{janowiecki2019} UDGs with resolved
\HI{} observations from \cite{gault2021} are disjoint from the subset of
\cite{janowiecki2019} UDGs in this work, we can do a simple test of our
\HI{} profile recovery alone using the \cite{gault2021} galaxies. }

\begin{figure*}[htb]
\centering     
\includegraphics[width=.95\linewidth]{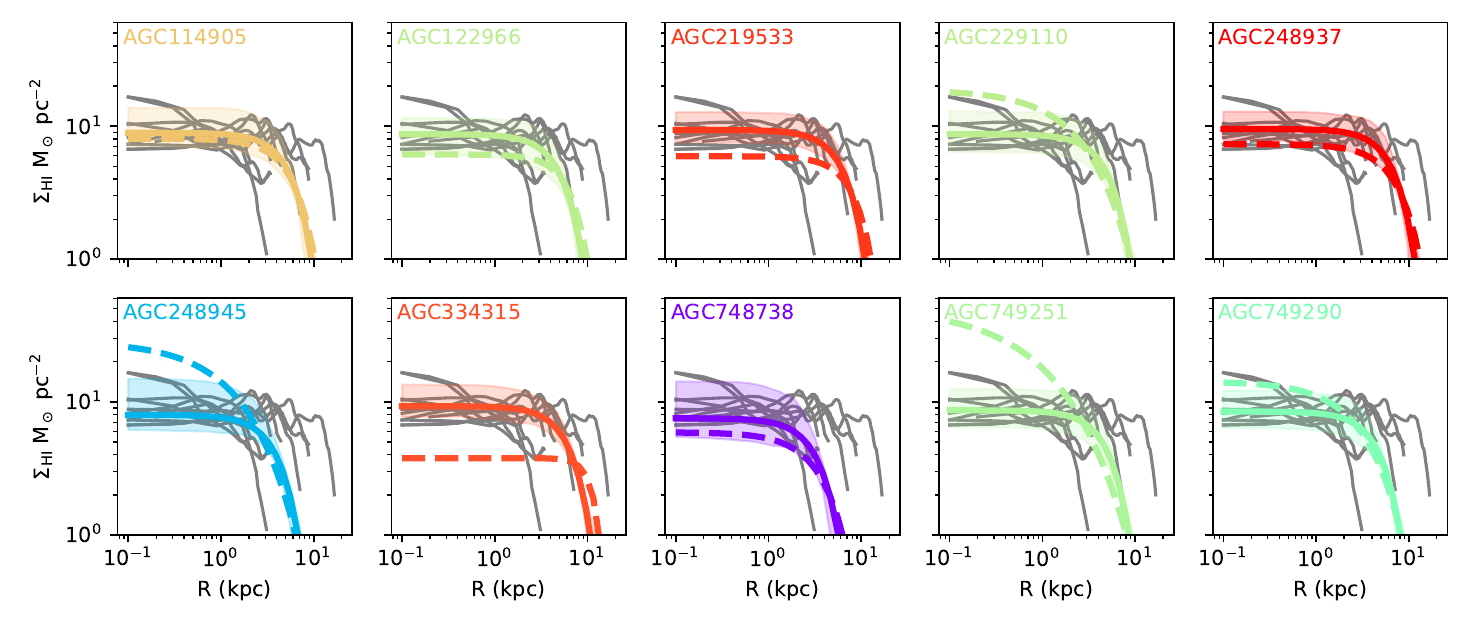}
\caption{ 
    \rrrtwo{
    In each panel we show the \HI{} radial profile inferred from the
    \cite{gault2021} reported structural parameters as a dashed curve and 
    our estimated radial profile as a solid curve. We also show the 
    directly measured profiles of \cite{leroy2008} as grey curves in each
    panel. We find that our estimation method adequately recovers the 
    UDG inferred radial \HI{} profiles, with significantly better performance 
    for inferred profiles close to the average profile -- an unsurprising result
    given that we estimate profiles by assigning galaxies average 
    \sersic{} parameters for their \HI{} mass.
    }
    } 
\label{f:gault_profiles}
\end{figure*}

\rrrtwo{In \autoref{f:gault_profiles} we show a comparison between 
the \HI{} profiles estimated from the \cite{gault2021} structural measurements
(dashed curves) with profile estimates analogous to those presented in the main
body of this work (solid curve and filled region). We also show the directly
measured profiles of \cite{leroy2008} as solid grey curves in each panel.
We find that the profiles are generally well-recovered in the range
of $\Sigma_{\rm HI}$ probed by the \cite{leroy2008} sample. We do find that
the inferred \cite{gault2021} profiles with very high \HI{} surface densities
(AGC749251, $\Sigma_{\rm HI} \gtrsim 30$ M$_\odot$ pc$^{-2}$) are significantly 
underestimated. However, we caution that the \cite{gault2021} profiles are 
estimated from their measured \HI{} masses, isophotal sizes, and 
fraction of \HI{} contained within that isophotal size -- not directly measured
profiles.
We also note that it is expected
theoretically that the maximum \HI{} surface density of these systems should be 
$\sim 10$ M$_\odot$ pc$^{-2}$ on kiloparsec scales \citep{krumholz2013}. 
}

\rrrtwo{We thus find that our average \HI{} profile estimation technique
works well to capture the average behavior of the \cite{gault2021} resolved
UDG measurements, though we caution that this test is quite limited in 
scope due to the inclusion of the \cite{gault2021} inferred profiles in 
creating the profile estimation method. 
As expected, the average estimation method performs better
for galaxies with average \HI{} surface densities than for UDGs with 
notably high or low \HI{} surface densities. Because our analysis focuses on
the bulk behavior of the dwarfs in our sample as opposed to the properties of
individual galaxies (and because high resolution \HI{} data are not yet available
for sufficiently large samples of UDGs), we find in this limited test case that
our estimation method is well-suited for the purpose of this work.}

\end{appendix}

\end{document}